\documentclass[iop,numberedappendix]{emulateapj}
\newcommand{\subsubsubsection}{\paragraph}

\usepackage{graphicx}
\usepackage{bm}
\usepackage{psfrag}
\usepackage{mathrsfs}
\usepackage{amssymb,amsmath}

\newcommand{\be}{\begin{eqnarray}}
\newcommand{\ee}{\end{eqnarray}}

\begin{document}






\title{An Asteroid Belt Interpretation for the  Timing Variations of the Millisecond Pulsar B1937+21}
\shorttitle{Circumpulsar Asteroids}

\author{R.~M.~Shannon\altaffilmark{1,2}, J.~M.~Cordes\altaffilmark{1}, T.~S.~Metcalfe\altaffilmark{3}, T.~J.~W.~Lazio\altaffilmark{4}, I.~Cognard\altaffilmark{5,6}, G.~Desvignes\altaffilmark{5,6,7}, G.~H.~Janssen\altaffilmark{8}, A.~Jessner\altaffilmark{7}, M.~Kramer\altaffilmark{7,8}, K.~Lazaridis\altaffilmark{7},   M.~B.~Purver\altaffilmark{8}, B.~W.~Stappers\altaffilmark{8},~\&~G.~Theureau\altaffilmark{5,6} }


\altaffiltext{1}{Astronomy Department, Cornell University, Ithaca,  New York, 14853, USA}
\altaffiltext{2}{Current Address:  CSIRO Astronomy and Space Science, Epping, NSW, 1710, Australia}
\altaffiltext{3}{Space Science Institute, 4750 Walnut Street Suite 205, Boulder CO 80301}
\altaffiltext{4}{Jet Propulsion Laboratory, California Institute of Technology, M/S 138-308, Pasadena, California, 91109,  USA}
\altaffiltext{5}{LPC2E/CNRS-Universit\'e d'Orl\'eans, Orl\'eans, F-45071, Cedex 2,France}
\altaffiltext{6}{Station de Radioastronomie de Nan\c{c}ay - Paris Observatory, France}
\altaffiltext{7}{Max-Planck-Institut f\"ur Radioastonomie, Bonn, D-53121, Germany}
\altaffiltext{8}{University of Manchester and Jodrell Bank Observatory, Manchester,M13 9PL, UK}
\shortauthors{Shannon~et~al.}
\email{ryan.shannon@csiro.au}
\email{cordes@astro.cornell.edu}

\begin{abstract}

Pulsar timing observations have revealed companions to neutron stars that include other neutron stars, white dwarfs, main-sequence stars, and planets.  
We demonstrate that the correlated and apparently stochastic residual times of arrival from the millisecond pulsar B1937$+$21 are consistent with the signature of an asteroid belt having a total mass $\lesssim 0.05~M_\earth$.  
Unlike the solar system's asteroid belt, the best fit pulsar asteroid belt extends over a wide range of radii, consistent with the absence of any shepherding companions.   We suggest that any pulsar that has undergone accretion-driven spin-up and subsequently evaporated its companion may harbor orbiting asteroid mass objects.  
The resulting timing variations  may fundamentally limit the timing precision of some of the other millisecond pulsars.
Observational tests of the asteroid belt model include identifying periodicities from individual  asteroids, which are difficult; 
testing for statistical stationarity that  become possible when observations are conducted over a longer observing span; and searching for reflected radio emission. 

\end{abstract}


\section{Introduction}

Arrival time measurements of rotation-powered  pulsars show time-correlated excesses after accounting for contributions from the spin-down of the pulsar, astrometric variations, and obvious orbital motion  \cite[][]{1985ApJS...59..343C}, often referred to as timing noise (TN), 
but which we will refer to as red noise (RN). 
In slowly spinning  canonical pulsars (CPs, pulsars with spin frequencies $ 0.5 \lesssim \nu \lesssim 30$~Hz, and relatively high surface magnetic fields $B \sim 10^{12}$~G), the excess is likely caused by rotational irregularities associated with a combination of magnetospheric torque fluctuations \cite[][]{2006Sci...312..549K, 2010Sci...329..408L} and instabilities arising from differential rotation of the neutron star (NS) crust and core \cite[][]{1990MNRAS.246..364J}.

Pulsars that have undergone accretion-driven spin-up and magnetic field attenuation are much more stable.  
The stability of these millisecond pulsars (MSPs,  $30 \lesssim \nu \lesssim 700$~Hz) has led to the discovery of the first Earth-mass planets outside of the solar system \cite[][]{1992Natur.355..145W}, stringent tests of general relativity through the study of gravitational interactions with white dwarf companions \cite[][]{2012arXiv1205.1450F}, and constraints on nuclear equations of state \cite[][]{2010Natur.467.1081D}.  
If timing precision can be improved further it may be possible to detect the signature of gravitational waves passing  through the solar neighborhood using an ensemble of pulsars \cite[i.e., pulsar timing array,][]{1990ApJ...361..300F}.

\begin{figure}[!ht]
\begin{center} 
\includegraphics[angle=-90,scale=0.5]{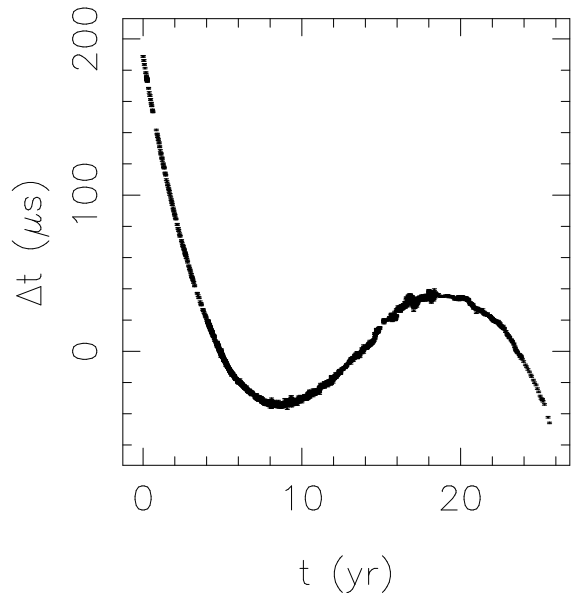} 
\caption{ \footnotesize \label{fig:TN_B1937} Residual TOAs $\Delta t$   for PSR B1937$+$21 over the $26$~yr observing span used in this analysis.}  
\end{center}
\end{figure}

 In some pulsars, the timing residuals contain temporally correlated stochastic noise that often appears consistent with  
 a `red' power spectrum that has significant power at low fluctuation frequencies.
 The most striking example of this in an MSP  is seen in the timing analysis of PSR B1937$+$21.  
In Figure \ref{fig:TN_B1937}, the post-fit residual times of arrival (TOAs) spanning $26$~yr are displayed for this object.    
The observations used to form these residuals came from the Arecibo, Effelsberg, Nan\c{c}ay, and Westerbork telescopes.  The observations and the model used to derive the residual TOAs are further described in  Appendix \ref{sec:app_1937_obs}.  
The residuals show correlated variations that are  similar in spectral nature to that observed in CPs, but at a much smaller level.   
The root mean square (rms) strength of the RN for the observations presented in Figure \ref{fig:TN_B1937} is $\sigma_{{\rm RN},2}\approx45~\mu$s, where the subscript `$2$' implies the rms residual after a second-order polynomial fit. 
In contrast, the rms strength of RN observed in many CPs can exceed $1$~s over a similar observing span.

This excess can be interpreted in a number of ways.  
The residuals could be associated with rotational instabilities similar to those that cause RN in CPs that  we will refer to as {\em red noise}.  
Indeed, a self consistent scaling relationship relating RN to spin frequency and  frequency derivative has  been developed that explains the measured levels and upper limits of RN in MSPs and the slower spinning CPs \cite[][]{sc2010}, which suggests that RN may be associated with the spin properties of the NS.
 

Additional contributions arise from refraction and scattering of radio waves in the interstellar medium (ISM).  However, these effects are highly wavelength-dependent.  After correction the timing residuals of PSR B1937$+$21 show a persistent achromatic component \cite[][]{1990ApJ...349..245C}, that we verify here.
 Achromatic effects arise from 
uncertainties in the solar system ephemeris used to transform topocentric TOAs to the solar system barycenter  \cite[][]{2010arXiv1008.3607C} and from the perturbations from
long-period gravitational waves \cite[][]{1979ApJ...234.1100D}, but these
effects are small and are correlated between pulsars.   The residuals seen in PSR~B1937+21
are much larger than those seen in other MSPs so these effects are secondary if not negligible.

\begin{figure}[!ht]
\begin{center} 
\includegraphics[angle=-90,scale=0.5]{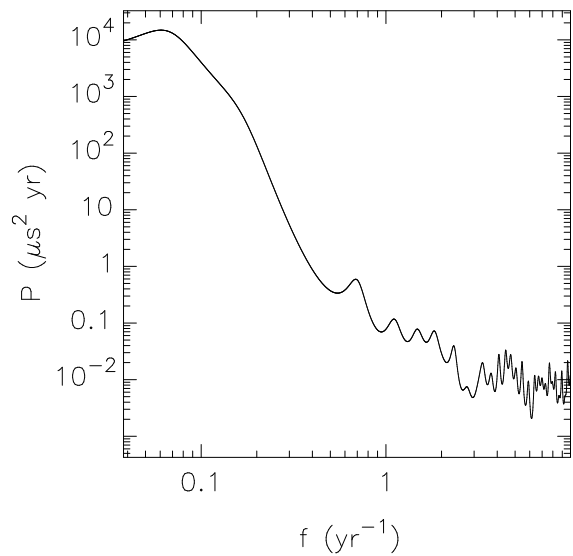}
\caption{ \footnotesize \label{fig:PS_B1937} Maximum entropy power spectrum for the observations of PSR~B1937$+$21 presented in Figure \ref{fig:TN_B1937}.  The methods for generating power spectra are discussed in Appendix \ref{app:maxent}.    The plateau in power at $f  < 0.1$~yr$^{-1}$ is associated with quadratic fitting introduced in TOA modeling, which high-pass filters the time series.   
}  \end{center}
\end{figure}

Here we investigate an alternative  mechanism for the correlated microsecond timing residuals in MSPs:  physical displacement of the pulsar due to recoil from circumpulsar objects.
While the residuals from PSR~B1937$+$21 are too small and too incoherent to be associated with a few large planets in compact orbits, they could be associated with a larger ensemble of asteroid-mass bodies that collectively produce the residuals.

  There is direct evidence that planetary and protoplanetary environments exist around some pulsars. 
Most obviously, there is the planetary system around the MSP~B1257$+$12 \cite[][]{1992Natur.355..145W,2003ApJ...591L.147K} comprising three moon to Earth-mass planets in $\approx 25$~d, $66$~d, and $98$~day orbits that induce an rms  scatter of $\approx 1$~ms on the residual TOAs.
There is also evidence for a planet in a wide orbit around the globular cluster pulsar B1620$-$26 and its binary white dwarf companion,   inferred from  both secular changes in the spin down of the pulsar and evolution of the WD-pulsar orbit \cite[]{1993Natur.365..817B,1999ApJ...523..763T}.  This planet was likely captured in a three-body interaction common in the  dense cluster environment \cite[][]{1993ApJ...415L..43S} representing a formation channel that will not be discussed further here.   

There is evidence for disks and planetary systems around other neutron stars as well. 
The young magnetar 4U~0142$+$61 shows excess mid-infrared  flux that can be attributed to thermal emission from a dust disk \cite[][]{2006Natur.440..772W}.  However,  the excess has also been attributed to  magnetospheric emission \cite[][]{2007ApJ...657..967B}.  
More recently, the impact of an asteroid-mass object on a neutron star has been proposed as the cause of an unusual gamma-ray burst \cite[][]{2011arXiv1112.0018C}. 

Circumstantial evidence comes from the likely presence of asteroid belts around another class of compact objects:  white dwarf stars (WDs). 
Accretion of asteroids is thought to pollute atmospheres and cause anomalously high metallicity in some WDs \cite[][]{2006A&A...453.1051K,2011arXiv1101.0158F}.
Tidal disruption of asteroids is hypothesized to cause dust disks around other WDs  \cite[][]{2002ApJ...572..556D}.   
While the planetary nebula phase that precedes the white dwarf end state for low mass stars is not as traumatic as the supernova explosion that precedes the neutron star end state for higher mass stars, the presence of asteroids around WDs suggests that rocky bodies can exist around post main sequence stars in relatively extreme environments.

We report our findings as follows:
in Section \ref{sec:residuals}, we discuss the perturbations to pulsar TOAs caused by asteroid belts.
in Section \ref{sec:formation}, we discuss the formation of asteroidal disks around pulsars.
in Section \ref{sec:stability}, we show that many configurations of asteroid belts can be stable for the entire many~Gyr lifetime of an MSP system;
in Section \ref{sec:disk_config}, we demonstrate through simulation that there are many plausible disk configurations that 
explain the observed residual TOAs of PSR~B1937$+$21;
in Section \ref{sec:comparison}, we compare the PSR~B1937$+$21 system to other sites of planet formation;
in Section \ref{sec:test_belt_model}, we motivate further observations that can be used to assess the plausibility of the asteroid belt model; 
and in Section \ref{sec:implications_timing_precision},  we discuss implications for precision pulsar timing.

\newpage
\section{Timing Perturbations from Circumpulsar Objects} \label{sec:residuals}

The reflex motion of a pulsar due to a companion causes pulse TOAs to vary. 
This effect has been  used to detect companion objects ranging from massive main sequence stars \cite[][]{1992ApJ...387L..37J}  to Earth mass planets \cite[][]{1992Natur.355..145W}.  

In this section, we first estimate the rms TOA  perturbation $\sigma_{\rm TOA}$ 
produced
 by an orbiting body
for comparison with the published rms residual, which enables us to make order-of-magnitude assessments of the size and structure of the system surrounding PSR~B1937$+$21. 
In later sections, we will compare the power spectrum of the observed residuals to those induced by simulated asteroid belts.   
 

For a single object with a mass $m$ orbiting a neutron star of 
mass $M_{\rm NS}$ in a circular orbit with a radius $a$, the rms 
residual TOA perturbation  after a second order fit 
 is \cite[][]{1993ASPC...36...43C}\footnote{Our 
expression differs by a factor of 
$1/\sqrt{2}$ from that given in \cite{1993ASPC...36...43C}  only because we present the rms residual rather than
the peak-to-peak value, and we include the transmission factor $F$. 
}
\be
\label{eqn:pert_single_object}
\sigma_{2}  &&\approx  F(P_{\rm orb},T) \left(\frac{a \sin i}{c\sqrt{2}} \right) \left( \frac{m}{M_{\rm NS}}\right) \nonumber \\
 &&\approx 0.85~{\rm ms}~  F(P_{\rm orb},T) \left(\frac{m}{M_\earth} \right)  \left(\frac{P_{\rm orb}}{1~{\rm yr}}\right)^{2/3} \sin i, 
\ee
where $i$ is the inclination angle of the orbit with respect to the line of sight. 
For a given mass, objects at larger radii produce larger timing perturbations.
 In the second line we have expressed  
the perturbation as a function of orbital period.
 
\begin{figure}[!ht]
\begin{center} 
\includegraphics[scale=0.5]{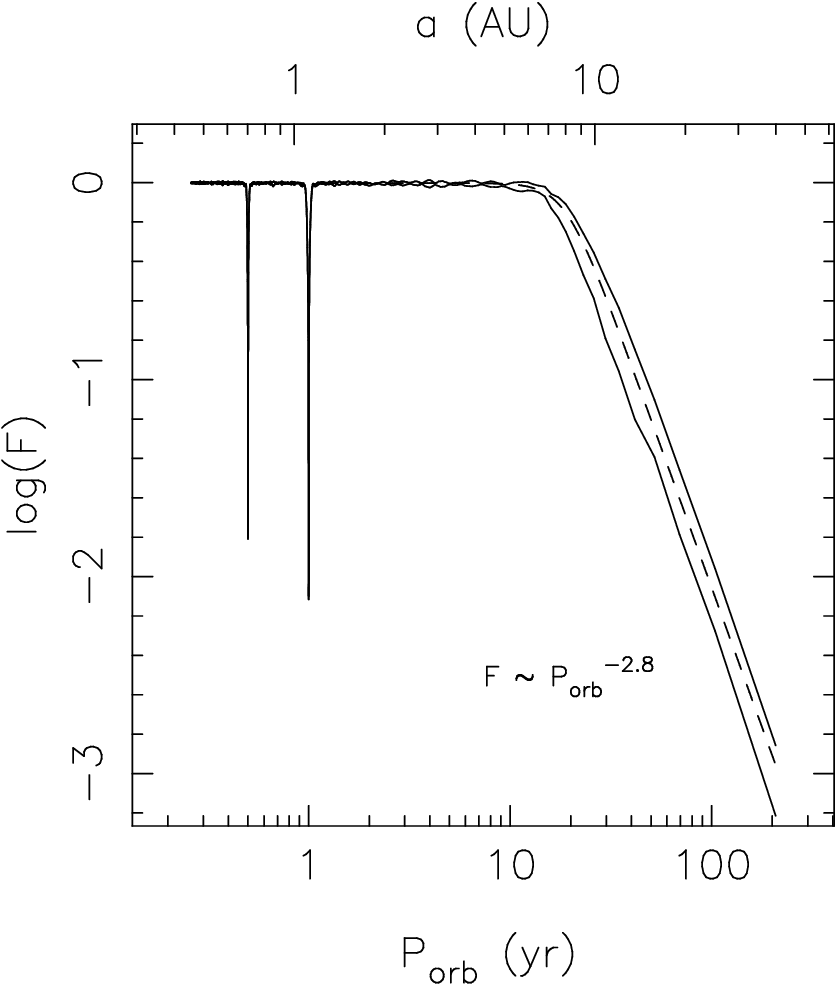}
\caption{ \footnotesize \label{fig:transfunc} Transmission function as a function of $P_{\rm orb}$, and semimajor axis $a$, derived from a set simulations described in Appendix \ref{app:transfunc}.
The solid lines show the $\pm 1$-$\sigma$ standard deviations about the mean (not displayed).  The dashed line shows the empirical relationship that we have adopted in Equation (\ref{eqn:trans_func_relationship}).   }  \end{center}
\end{figure}

  The transmission function 
   $F(P_{\rm orb},T)$ accounts for the TOA fitting procedure  
used to model pulsar spin phase,  frequency and frequency derivative, 
which removes any parabolic perturbation to pulsar TOAs.   
In Figure \ref{fig:transfunc} we show $F(P_{\rm orb},T)$ for our observing cadence, derived through simulations described in Appendix~\ref{app:transfunc}. 
  Ignoring the astrometric terms (which absorb power in two  narrow windows at orbital periods of $1/2$ and 1~yr, which are unimportant to this analysis), we find an empirical relation 
\be
\label{eqn:trans_func_relationship}
F(P_{\rm orb}) = \frac{(P_{\rm orb}/19~{\rm yr})^{-2.83}}{\sqrt{1+(P_{\rm orb}/19~{\rm yr})^{-5.67}} },
\ee
which is displayed as the dashed line in Figure \ref{fig:transfunc}.
We find that when $P_{\rm orb} \ll 19$~yr, $F\approx 1$ and when   
$P_{\rm orb} \gg 19$~yr, $F\propto P^{-2.83} \propto a^{-4.25}$.
The effect of the outer regions on the disk is suppressed by the necessary quadratic fitting.  
Nonetheless, the outer regions can have significant influence on the residual TOAs if the 
objects have sufficiently large mass.

For PSR~B1937$+$21, the lack of an obvious periodicity implies that there is no single, sufficiently massive companion with a period shorter than approximately $26$~yr, which is the observing span of our observations.   The total mass in a large ensemble of asteroids at random orbital phases can exceed that of a single planet by several orders of magnitude because the residuals are an incoherent sum of the individual sinusoidal contributions.  
For $j=1,N_a$ asteroids, the rms perturbation to the TOAs is

\be
\sigma_{\rm TOA} \approx \left[  \sum_{j=1}^{N_a} \sigma_{2,j}^2\right]^{1/2}
\ee

If all asteroids have the same inclination, as would be the case if they originate from a disk, and we make the simplifying assumption that the masses and semi-major axes of the asteroids are statistically independent,   the rms of the time series can be expressed as
\be
\label{eqn:rms_ts}
\sigma_{\rm TOA} &&\approx \left( \frac{\sin i}{c M_{\rm NS} } \right)  \left(\frac{N_a \langle a^2 F^2 \rangle \langle m^2 \rangle}{2}\right)^{1/2}, \nonumber \\
 && \approx 0.76~{\rm ms}~ \sin i \left( \frac{N_a}{100}\right)^{1/2} \left( \frac{\langle a^2 F^2\rangle^{1/2}}{10~{\rm AU}}  \right) \left(\frac{\langle m^2\rangle^{1/2}}{10^{-2} M_\earth} \right), 
\ee
where
\be
\langle a^2 F^2 \rangle &&= \int_{a_{\rm min}}^{a_{\rm max}} a^2 F^2(P(a), T) f_A(a) da,\\
\langle m^2 \rangle &&= \int_{M_{\rm min}}^{M_{\rm max}} m^2 f_M(m) dm, 
\ee
  $f_{A}$ is the   probability density function (PDF) for $a$, and $f_{M}$ is the PDF for $m$,.

By defining the total mass of the asteriod belt in terms of the mean asteroid mass,
 $M_{\rm belt} = N_a \langle m \rangle$, we express the rms residual  as 
\be
\label{eqn:total_pert}
\sigma_{\rm TOA} \approx 
0.76~{\rm ms}~ 
\zeta \sin i
\left( \frac{N_a}{100} \right)^{-1/2}
\left(\frac{\langle a^2 F^2 \rangle^{1/2}}{10~{\rm AU}} \right) 
\left(\frac{ M_{\rm belt}}{1~M_\earth}\right), 
\ee
where $\zeta$ is the ratio of the rms to mean mass:
\be 
\zeta \equiv \frac{\langle m^2 \rangle^{1/2}}{\langle m \rangle}.
\ee
If the mass probability density function follows a power-law, i.e. $f_M(m) \propto m^{-\alpha}$, in the mass range $M_{\rm min}$ to $M_{\rm max} \gg M_{\rm min}$,
\begin{equation}
\zeta = \left\{ 
\begin{split}
& \frac{(2-\alpha)}{\sqrt{(3-\alpha)(1-\alpha)}}~~~~~~~~~~~~\alpha < 1 \\
& \frac{(2-\alpha)}{\sqrt{(3-\alpha)(\alpha-1)}} \left( \frac{M_{\rm max}}{M_{\min}}\right)^{(\alpha-1)/2}~~1< \alpha < 2, 
\end{split}
\right.
\end{equation}

The values of $\zeta$ show marked variation depending on the configuration. 
If the disk has a relatively flat mass distribution  or a narrow mass distribution 
with $M_{\rm min} \approx M_{\rm max}$, then $\zeta \approx 1$.  
A narrow mass distribution  could arise if a dynamically cold disk of planetesimals was formed after a runaway growth phase, a scenario we discuss below. 

It is possible that the objects have a wide range of masses with a steep distribution.
If the system is supported by collisional interactions,  $\alpha = 11/6$ \cite[][]{1969JGR....74.2531D} and objects could have a wide range of masses. 
Observations indicate that the size distribution of Solar system objects has a distribution that is somewhat less steep with $\alpha \approx 1.5$ \cite[][]{2011PASJ...63..335T}. 
In both of these cases $\zeta$ could be much larger than unity.
For example, $\zeta \approx 990$ if $\alpha = 11/6$, $M_{\rm max}  =  0.1 M_{\earth}$, and $M_{\rm min} = 10^{-10} M_{\earth}$.  This value of $M_{\rm min}$ corresponds to the size of the smallest object that can survive radiation driven migration to the current age of the pulsar, as described in the analysis in Section \ref{sec:rad_migration}
The value of $M_{\rm max}$ corresponds to the maximum size of an object that would remain undetectable in spectral analysis (see Section \ref{sec:detect_period}).

Using $\sigma_{\rm TOA} =  45~\mu s$ for  PSR~B1937$+$21, Equation
(\ref{eqn:total_pert}) constrains 
 disk parameters according to
\be 
\label{eqn:num_asteroids_2}
\zeta \sin i
 \left(\frac{N_a}{100}\right)^{-1/2} 
 \left(\frac{{\langle a^2 F^2\rangle}^{1/2} }{\rm 10~AU}\right)  
 \left(\frac{ M_{\rm belt}}{1~M_\earth}\right) 
 \approx 0.06,
\ee 
showing the clear trade off between the 
total mass in asteroids, the number of objects,  and their distribution in semi-major axis and mass.
For any configuration, the range of semi-major axes of the largest bodies can be constrained by matching the spectral content of the residuals to simulations.  



\section{Asteroid Formation}\label{sec:formation}

Several mechanisms have been proposed for the formation of planetary systems around pulsars.  
They include:  (1) formation along with the progenitor \cite[][]{1993ASPC...36..371P,1993ASPC...36..149P}, (2) formation from supernova fall back material \cite[][]{1991Natur.353..827L}; or (3) formation from material left over from the recycling phase \cite[][]{1991ApJ...382L..81N}.    
The second is probably relevant to the disk around the magnetar  4U~0142$+$61, and may also play a role in the variability of the radio emission observed in many pulsars \cite[][]{2008ApJ...682.1152C}.   
The third case is probably relevant to PSR~B1257$+$12.\footnote{For an alternative explanation, see \cite{2001ApJ...550..863M}.}
In this scenario a lower mass companion star evolves, over-flows its Roche lobe, and donates material to the neutron star, spinning it up from a radio-quiet 
object to an MSP with pulsed radio emission and a large particle wind. 

After  (or perhaps during) recycling, the particle wind can ablate its companion, as is the case in the PSR~B1957$+$20 system \cite[][]{1988Natur.333..237F} and a growing number of similar pulsars.  
A number of the MSPs recently discovered in radio searches of Fermi gamma-ray point sources  are ``black widow'' objects \cite[][]{2011AIPC.1357..127R} in which the MSP is ablating its former mass-donor star. 
Additionally a growing number of MSPs are observed with low mass $\ll 0.1 M_\odot$ WD companions, with  the MSP PSR J1719$-$1438 having nearly ablated its WD companion, transforming a star that was initially a large fraction of a solar mass into an object with a mass of $10^{-3}~M_\odot$ \cite[][]{2011arXiv1108.5201B}.



The disk formation commences when the MSP disrupts the companion. 
While most of the material from the disrupted companion is accreted from the disk into the pulsar,  conservation of angular momentum requires dispersal of a fraction of the material  over a wide region around the pulsar. 
The disk subsequently  expands and cools enough to form the  dust particles that seed the formation of larger objects.  At this early time the disk may contain $ > 10~M_\earth$ of dusty material extended over many AU (i.e. much more than in the present-day system, as we
show below), well beyond the tidal disruption radius of the pulsar.    
High angular momentum systems  may extend to $\approx 20$~AU \cite[][]{2007ApJ...666.1232C}.



The high luminosity of PSR~B1937$+$21 would prevent the formation of planetesimals and the survival of asteroids in the inner region of any disk.
To estimate the radius where this happens,  we assume thermal equilibrium of the material with input flux  $F_{\rm in}$, and that the disk is optically thin.  
The input flux is dominated by the pulsar spin down luminosity  $L= 4\pi^2 I \nu \dot{\nu} \approx 250 L_\sun$ where $I$ is the neutron star moment of inertia and is assumed to be $10^{45}~{\rm g}~{\rm cm}^{3}$. 
 The pulsar spin down luminosity comprises low-frequency Poynting flux, a wind of relativistic particles, and high energy magneotspheric radiation.
The pulsar also emits thermal radiation that in the case of PSR~B1937$+$21 is negligible compared to this spin down luminosity ($L_{\rm thermal}  = 10^{-4.7}  L_{\odot} (R_{\rm NS}/{\rm 10~km})^{2} (T/10^5 {\rm K})^{4}$, where $R_{\rm NS}$ is the neutron star radius and $T$ is its temperature).    

The flux incident on an object at orbital radius $r$ is
\be
\label{eqn:fin}
 F_{\rm in } = \frac{\epsilon g_b L}{4 \pi r^2},
 \ee
 where $\epsilon$ is the fraction of the total spin down loss rate that
can be absorbed and $g_b$ is a factor that accounts for beaming of the pulsar's energy
loss. 
The heating of the material depends on the fraction of the luminosity beamed toward the disk.  In many models of pulsar emission, the spin-down energy is beamed out of the open field line region around the magnetic poles \cite[][]{1969ApJ...157..869G}.   However, recent numerical simulations have suggested that the beaming is closer to isotropic \cite[][]{2011heep.conf..139S}.
For the rest of the paper present scaling laws based on the fiducial values $\epsilon =1$ and $g_b =1$.

Assuming an asteroid radiates like a blackbody, the radiated flux is
\be
\label{eqn:fout}
F_{\rm out}=\sigma_{\rm SB} T^4,
\ee
where $\sigma_{\rm SB}$ is the Stefan-Boltzmann constant. 
If the asteroid is in thermal equilibrium (i.e., $\pi R_a^2 F_{\rm in} = 4 \pi R_a^2 F_{\rm out}$ if the asteroid is spherical), the equilibrium temperature $T$ is

\be
\label{eqn:disk_temp_radius}
T(r) = 1100~{\rm K}~\left(g_b\epsilon \right)^{1/4} \left(\frac{L}{250 L_\sun}\right)^{1/4}  \left(\frac{r}{\rm AU} \right)^{-1/2},
\ee 
suggesting that rocky material (melting point $T_{\rm melt} \approx 1100$~K) can survive at $r \gtrsim 1$~AU and a more metal-rich
object (melting point $\approx 1500$~K) could survive to about $1/3$ of this distance. 

The formation of rocky macroscopic objects follows a  path similar to that of  planet formation in main sequence stellar systems and planet formation in the PSR~B1257$+$12 system \cite[][]{2009ApJ...691..382H}, where
 protoplanets undergo runaway growth until they reach a mass limited by their ability to 
gravitationally attract material from the surrounding 
surface mass density \cite[][]{1987Icar...69..249L}.

Slower but inexorable growth of objects to planet sizes follows if perturbations induce orbit crossings.   
The rate of growth in this secondary phase depends on protoplanet masses, velocity dispersion, and the presence of gas to induce orbit crossings and accrete onto more massive planets. 

Our proposal is that 
the  evolution of the PSR~B1937$+$21  differs from other known systems  because its progenitor disk was too tenuous to form Earth-mass  or larger planets.

The largest isolation mass (the Hill mass) $m_H$ to which a protoplanet may grow during the runaway growth  phase is the total mass contained within an orbital radial range of several Hill radii \cite[][Equation 11]{1987Icar...69..249L}:
\be
m_H = \frac{(4 \pi B r^2 \Sigma)^{3/2}}{(3 M_{\rm NS})^{1/2}} \approx 10^{-2.8} M_\earth \left(r_{\rm AU}^2 \Sigma_{0} \right)^{3/2},
\ee 
where  $B\approx 3.5$ is a numerical constant and $r_{\rm AU}$ is the semi-major axis in AU, $\Sigma$ is the surface mass density,  and $\Sigma_{0}$ is the surface mass density in g~cm$^{-2}$.
Without any means for inducing orbit crossing,  
objects do not feed well beyond their Hill spheres, resulting in a disk comprised of asteroid-mass objects.

Planetesimals or protoplanets formed in such a disk would be too small to force gravitational interaction after an initial runaway growth phase, and the disk would comprise many low-mass bodies in marginally stable orbits.  
The disk would extend over a wider range of orbital radii than the solar system's asteroid belt because there are no planet-mass bodies to impose radial confinement.

We suggest that disks of planetesimals may be a generic outcome of accretion driven spin-up and 
may potentially provide a noise floor for timing precision.  The level of this floor would depend on the disk configuration, which itself evolves over the Gyr lifetime of the MSP.

The surface density at the time of planetesimal formation may have been larger if volatiles were present, and the Hill mass may be correlated with orbital radius and depend on the radial variation of $\Sigma$.
Even in this case, our basic argument is unchanged.

\section{Disk Stability}\label{sec:stability}

In this section, we investigate the stability of disks of asteroids to gravitational and radiative migration and show that these disks can persist to the present age of the pulsar. 
The spindown age of PSR B1937+21 is $\nu/2 \vert\dot{\nu}\vert \approx 250$~Myr and is an upper bound on the elapsed time since the cessation of accretion 
\citep[e.g.][]{1999ApJ...520..696A,2000ApJ...528..401W}. 
\cite{2012Sci...335..561T} includes a discussion on pre-accretion time scales and also suggests that spin down age likely overestimates the post-accretion age the system.   
Even in this latter case, the age is much longer than the expected runaway growth phase, but is short enough that the asteroidal disk we associate with timing residuals may still be evolving: slow coalescence may occur while the wind evaporates the protoplanets that migrate to orbits near the NS \cite[][]{2008ApJ...682.1152C}.
While direct collisions between disk objects are rare, 
gravitational stirring induced by a mass dispersion in the objects introduces a viscosity that will affect the lifetime of a massive system.
Pulsar radiation  clears the disk of its smallest objects.

\subsection{Gravitational Migration}

The arguments presented here follow those presented in \cite{2010MNRAS.401..867H}, which examines the stability of  planetesimals in a 
dynamically cold disk in which orbits do not cross. 

We first assume that all objects have a characteristic mass $m$, located in a disk of surface density $\Sigma$.   
In this case, the typical separation between objects is
\be
\label{eqn:radial_spacing}
\Delta r = \frac{m}{2 \pi \Sigma r}.
\ee

\cite{1996Icar..119..261C} examined the stability of planetary systems of up to $20$ objects.  They found that systems containing $20$ objects became unstable after $\approx 10^8$~Myr  if the orbital separation satisfies
 \be
 \label{eqn:hill_criterion}
 \Delta r  \approx 10  r_H.
 \ee
 where
\be
    \label{eqn:hill_radius}
   r_H = r \left( \frac{m}{3M_{\rm NS}}\right)^{1/3}
  \ee 
is the Hill radius.

\cite{1996Icar..119..261C} also found that the number of planets and the mass dispersion of the planets does not affect the stability time of the systems and    
argue that systems with orbital separations  larger than $10~r_H$ are likely stable because 
the time scale for interactions increases exponentially with distance. 
We assume that the results of \cite{1996Icar..119..261C} can be extended to systems with a larger numbers of objects, and that objects spaced by $10~r_H$ can persist for at least~$250$~Myr.


For a belt of objects of mass $m$ located between $r_1$ and $r_2$ the number of objects is therefore limited to 
\be
N_a  \lesssim \int_{r_1}^{r_2} \frac{dr}{\Delta r} = \frac{1}{10} \ln\left(\frac{r_2}{r_1} \right) \left( \frac{3 M_{\rm NS}}{m}\right)^{1/3}. 
\ee
Substituting $m \rightarrow \langle m \rangle =  M_{\rm belt}/N_a$ and solving for $N_a$, we find
\be
 \label{eqn:num_asteroids}
N_a  && < \sqrt{\frac{3}{1000}}  \left[\ln \left(\frac{r_2}{r_1} \right)\right]^{3/2} \left( \frac{M_{\rm NS}}{M_{\rm belt}}\right)^{1/2} \nonumber \\
&& < 50 \left[\frac{\ln(r_2/r_1)}{\ln(10)}\right]^{3/2} \left( \frac{M_{\rm belt}}{M_\earth} \right)^{-1/2}.
\ee

We calculate the maximum-mass asteroid belt by using the maximum allowed number of asteroids combined with Equation~(\ref{eqn:total_pert}), which implies
$M_{\rm belt} \propto \sigma_{\rm TOA} N_a^{1/2}$.   Numerically we obtain
\be
M_{\rm belt, max} 
 &\approx&  
 	0.08 M_\earth 
	\zeta^{-4/5} 
	\left(\frac{\ln r_2/r_1}{\ln 10} \right)^{3/5}	
	\nonumber \\
	 \times &&
	\left( \frac{ \sin i \langle a^2 F^2  \rangle^{1/2} }{\rm 10~ AU}\right)^{-4/5} 
	\left( \frac{\sigma_{\rm TOA}}{45~\mu{\rm s}}\right)^{4/5}.
\ee


\subsection{Radiation Driven Migration} \label{sec:rad_migration}

Radiation forces will cause small objects to migrate through the system on time scales shorter than the age of PSR~B1937$+$21.
Here, we examine the effects of two different types of forces:  radiation drag (Poynting-Robertson drag and an analogous effect due to the relativistic pulsar wind) and Yarkovsky migration.

The Poynting-Robertson effect causes small particles to migrate from the Earth into the Sun 
on a time scale $t_{\rm PR} = 4 \pi \rho c^2 r^2 R/3L \approx 672~{\rm yr}~\rho R_{-4} r_{\rm AU}^2 L_{\odot}/L$ \cite[][]{1979Icar...40....1B}, where $\rho$ is the particle density in g~cm$^{-3}$, $R_{-4}$ is the grain size in  $10^{-4}$~cm, $r_{\rm AU}$ is the orbital radius in AU, and $L$ is the luminosity.    
For a $10~\mu$m grain orbiting PSR~B1937$+$21 at $1$~AU, the time for the grain to migrate to the pulsar is
\be
&&t_{\rm PR} \approx 2.7~{\rm yr} 
(g_b \epsilon)^{-1} \left(\frac{\rho}{\rm 1 g~cm^{-3}} \right) \left( \frac{R}{10^{-4} {\rm cm}} \right)
\nonumber \\
&&
\quad\quad\quad\quad
\times  \left( \frac{r}{\rm AU}\right)^{2} \left(\frac{L}{250 L_{\odot}}\right)^{-1}.
\ee
Grains migrate very quickly in any compact disk surrounding the pulsar and little dust is expected close to the pulsar, unless there is a large reservoir of material further away. 

For much larger objects, the Yarkovsky effect will cause migration over the lifetime of the pulsar system.
The diurnal Yarkovsky effect is the net force on an illuminated
object whose afternoon surface is hotter than its
nighttime surface. It can lead to an increase or decrease in
semi-major axis depending on the lag angle of the hottest
surface. In contrast, the seasonal Yarkovsky effect applies to objects
with spin axes tilted from their orbital plane, causing a net
force along the spin axis that always decreases the semimajor
orbital radius. 

Here we  consider only  the seasonal Yarkovsky effect, which yields
a radial migration rate 
\be
&&\dot{r}\approx   -5~{\rm AU~Gyr^{-1}} \nonumber \\
&&~~ \times g_b \epsilon~\left(\frac{r}{\rm 1 AU}\right)^{-1/2} \left(\frac{R_a}{\rm 1~km}\right)^{-1}  \left( \frac{L}{250~L_\odot}\right),
\ee
where we have re-scaled Equation (A4) of \cite{2008ApJ...682.1152C} using fiducial values
better matched to the  disk configuration under analysis here.  

We define a critical asteroid size $R_{\rm yar}$ such that all objects smaller than $R_{\rm yar}$  migrate inward from a distance $r$ to the pulsar magnetosphere\footnote{  Objects that have migrated inward will be evaporated well before they reach the magnetosphere
of the MSP (defined as the light cylinder radius, $r_{\rm LC} = cP/2\pi \approx 74$~km.) and 
thus are not expected to produce any intermittency effects or torque fluctuations
\cite[e.g.,][] {2008ApJ...682.1152C}.}  in a time $T$.   Scaling to PSR~B1937$+$21, we find,         \be
\label{eqn:yark_criterion}
R_{\rm yar} =  5.1~{\rm km}~g_b \epsilon \left( \frac{r}{\rm 1~AU} \right)^{-1/2} \left( \frac{L}{250 L_{\odot}}\right) \left(\frac{T}{250~{\rm Myr}}\right).
\ee

The Yarkovsky effect therefore can cause   objects up to $\sim 5$~km to migrate inward
over the Gyr lifetime of many MSPs and presumably be evaporated well outside the
light-cylinder radius. 
For spherical asteroids with densities of $1$~g~cm$^{-3}$, $R_{\rm yar}$ corresponds to objects with masses of $\approx 10^{-10}~M_{\earth}$.
The Yarkovsky effect cannot cause larger objects than this, or much more distant than $1$~AU,  to migrate appreciably over the lifetime of the system.

We conclude that a disk extending beyond $\approx 1$~AU, populated by objects with sizes larger than $\sim 5$~km would survive radiative effects up to the present day for
PSR~B1937$+$21 and thus could induce timing variations at the present epoch.

\section{Plausible Disk Configurations} \label{sec:disk_config}
 
 The residual TOAs for PSR~B1937$+$21 are dominated by low-frequency 
components that are
nevertheless spread over a fairly broad band. 
 To find asteroid systems that reproduce the frequency content of the data, simulations of a wide range of asteroid belt configurations were conducted.  
  Each asteroid was randomly assigned an orbital radius according to a probability distribution that is flat in radius (hence decreasing in asteroid surface density with radius) between an inner radius $r_1$ and an outer radius $r_2$.
Within its orbit, each asteroid was given a random initial phase.
   The masses of the objects were selected from a power law mass distribution.
   The asteroids were modeled to be non-interacting, which is an appropriate approximation because the evolution of the orbits over the $26$~yr observing span (corresponding to at most a few orbits) is small.    Additionally, all objects within the simulated disks were required to 
 satisfy the dynamical and thermal migration criteria defined in Equations (\ref{eqn:hill_criterion}) and (\ref{eqn:yark_criterion}), respectively.

Random placement of objects alone was found to be computationally inefficient at producing dynamically stable disks when the number of objects was greater than $\approx 500$,  because the probability of an object being placed within $10~r_H$ of another object is large.
 To increase computational efficiency,  we edited out objects that were in dynamically unstable orbits.  Disks were iteratively populated disks with an increasing numbers of objects. If any object was initially placed in an orbit within $10$ Hill radii of another object it would be deleted from the disk.  Consequently, simulated disks contained fewer objects than initially specified.    
 We found that it was also necessary to edit simulations that contained detectable individual sources.  The minimum detectable asteroid mass is discussed in Section \ref{sec:detect_period}.  Based on analysis of that section, we restricted objects to have masses $m < 5\times 10^{-4} M_\earth (P/1~{\rm yr})^{2}$ for $f > 1/26$~yr$^{-1}$.

 At each epoch (corresponding to the observation times in the actual data), the cumulative orbital effect from all asteroids was calculated.   To this, a random white-noise contribution of rms amplitude equal to the formal TOA uncertainty  was added. 
The residual TOAs were found by subtracting from the simulated TOAs a least-squares fit comprising quadratic terms that model pulsar spin down.
To verify that the simulation code worked, we generated systems comprising one or a few objects with known masses.  
We found that the amplitude of the TOA perturbations agreed with the expressions in Equation (\ref{eqn:pert_single_object}).  

The real and simulated data were compared using maximum entropy power spectra of the timing residuals.
The quality of the match was assessed using a cost function,
\be
\label{eqn:cost_function}
C = \sum_k \left[\ln(S_k^{(o)}/S_k^{(s)})\right]^2, 
\ee
where $S_k^{(o)}$ and $S_k^{(s)}$ are the power spectra of the observed and simulated data sets, respectively.  
We searched for the minimum-cost combinations of parameters by using a brute-force search over a grid of values for parameters that are displayed in Table \ref{tab:grid_search_params}.
For each unique combination of parameters, $100$ stable asteroid belt configurations were simulated. 
We only simulated disks comprising  $N_a < 4000$ objects with a minimum mass much larger than the minimum mass object that can survive. 
For power law mass distributions with  $\alpha < 2$, the mass dispersion $\langle m^2 \rangle$, and hence $\sigma_{\rm TOA}$ (see Equation \ref{eqn:rms_ts})  are dominated by the largest objects, and (for our purposes) it is not necessary to simulate the smallest objects. 
This was done primarily to reduce the computational burden.

\begin{deluxetable*}{cll}
\tabletypesize{\footnotesize} \tablecolumns{3}
 \tablecaption{Grid search parameters\label{tab:grid_search_params}}
\tablehead{ \colhead{Parameter}  & \colhead{Description} &\colhead{Values}}   
\startdata

$M_{a}$ & Total mass ($M_\earth$) & $0.001$, $0.003$, $0.01$, $0.03$, $0.1$, $0.3$ \\
$N_a$ & Number of asteroids & $3$, $10$, $30$, $100$, $300$, $1000$, $2000$, $4000$ \\
$a_{\rm min}$ & Inner radius of asteroid belt (AU)& $0.5$, $1.0$,  $3.0$,  $10.0$,  $30.0$ \\
$a_{\rm max}$ & Outer radius of asteroid belt (AU) & $1.0$,  $3.0$, $10.0$, $30.0$, $100.0$ \\
$\alpha$ & Spectral index of mass distribution &  $0.1$, $0.5$, $0.9$, $1.1$, $1.25$, $1.5$, $1.67$, $1.83$ \\
$M_{\rm min}$ & Minimum mass of asteroid ($M_\earth$) & $10^{-8}$ \enddata
  \tablecomments{Grid Search Parameters.  For each unique set of parameters, $100$ stable asteroid belt configurations were simulated and compared to the observed residuals using the cost function expressed in Equation (\ref{eqn:cost_function}).  }
 \end{deluxetable*}

\begin{figure*}[!ht]
\begin{center} 
\begin{tabular}{c}
\includegraphics[angle=0,scale=1.0]{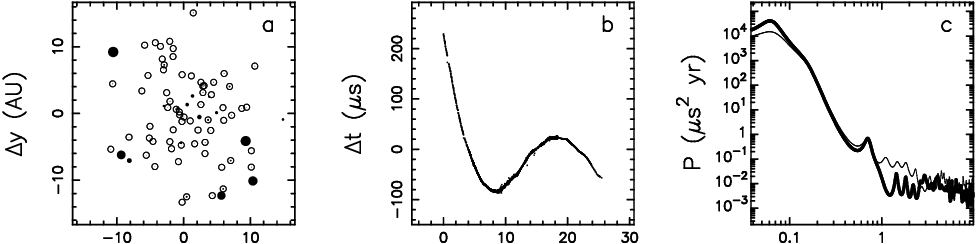} \\
\includegraphics[angle=0,scale=1.0]{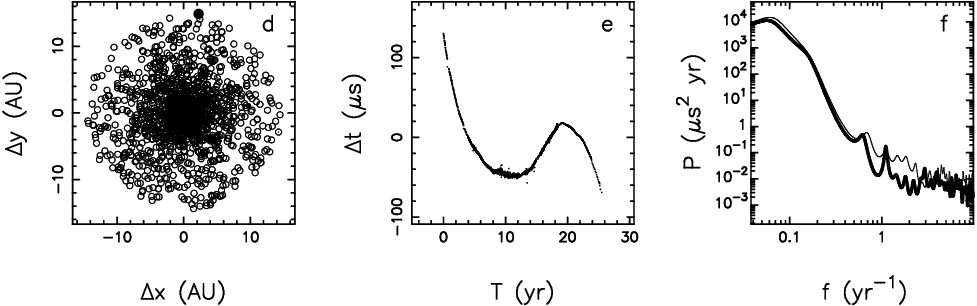}
\end{tabular}
\caption{ \footnotesize \label{fig:best_config} Best-fit asteroid belts and their signatures in residual TOAs.   In the top row (panels $a$-$c$)  we show a configuration comprising $80$ objects with a total mass of $0.04~M_\earth$.    In the bottom row (panels $d$-$f$) we show a configuration comprising $1609$ asteroids with a total mass of $0.06~M_\earth$. { \em Leftmost panels ($a$ and $d$): } Asteroid belt configuration at initial time in the simulation.  Each asteroid is represented by a filled circle, with the size of the circle proportional to the radius of the asteroid.  Open circles highlight objects with small masses.        {\em Center panels ($b$ and $e$)}: Residual TOAs associated with the asteroid belt.  {\em Rightmost panels ($c$ and $f$)}:  Maximum entropy power spectra for the simulated residuals (thick lines) and the observed time series (thin lines).       }  
\end{center}
\end{figure*}

 No unique combination of parameters was found that minimized the cost 
function. 
 This is highlighted by two of the best-matching asteroid belt configurations, which are displayed in 
 Figure \ref{fig:best_config}.  One corresponds to an asteroid belt with $N_a=80$ objects, total mass\footnote{The total disk masses do not match the  input {\em grid-search} values listed in Table \ref{tab:grid_search_params} for two reasons. Firstly, asteroids were randomly assigned masses, so agreement with the injected value is only expected in the ensemble average.   Variation about this mean is expected, especially in cases where the mass distribution has a steep negative spectral exponent.  Secondly,  as justified above, dynamically unstable objects were culled from systems, reducing disk masses. }     $M=0.04 M_\earth$, and a relatively flat mass distribution ($\alpha = 1.1$), 
 while the second has $1609$ objects, a total mass of $M=0.06 M_\earth$ with a steeper mass distribution ($\alpha = 1.5$). Both belts extend to an orbital radius of approximately $15$~AU and therefore contain objects orbiting at periods longer than the data span length. 

The number of asteroids can vary in the best-matching configurations from a few tens to a  few thousands, but the total asteroidal mass is found to approximately satisfy Equation (\ref{eqn:num_asteroids_2}).

\section{Comparison With other Planetary Systems}
\label{sec:comparison}

 \cite{2007ApJ...666.1232C} and \cite{2009ApJ...691..382H}  modeled 
the formation of solid objects in disks formed by either 
supernova fallback or tidal disruptions of companion white dwarf stars
with an emphasis on 
modeling the PSR~B1257$+$12 planetary system.  The predominant difference 
between  the two types of disks is their initial angular momentum, with the tidal disruption disk containing larger initial angular momentum.  
They found that both kinds of disks produced planets in 
many-AU orbits, much larger than the $0.2$~to~$0.5$~AU orbits of 
the planets around  PSR~B1257$+$12.   The tidal disruption planets were found to extend to orbital radii $> 20$~AU.   These disks compare favorably to our best-fit models which have objects orbiting at distances of $\sim 20$~AU. 

Comparison with our maximal-mass solution implies that at $1$~AU, the present surface density is $\Sigma \approx 0.013~{\rm g}~{\rm cm}^{-2}$, smaller by three orders of magnitude than the minimum-mass surface density (i.e., excluding volatiles) thought to be  present in the solar system at the end of the runaway phase \cite[][]{1993ASPC...36..217L}, and smaller by $\gtrsim 10^{4}$ than in the planet forming region near the planet pulsar system PSR~B1257$+$12 \cite[][]{1993ASPC...36..197R,2007ApJ...666.1232C,2009ApJ...691..382H}.  However, our estimate of the surface density may underestimate the initial surface density due to the removal of material caused by the effects discussed in Section \ref{sec:stability}.

\section{Tests of the Asteroid Belt  Model}\label{sec:test_belt_model}

A drawback of an asteroid belt cause for the RN observed in the timing residuals of  PSR~B1937$+$21 is that it is difficult, though not impossible, to test.  
In the following subsections we first highlight tests that can be conducted using high precision pulsar timing, and then discuss tests generic to detecting circumpulsar objects.

\subsection{Pulsar Timing Tests}

\subsubsection{Detecting the Periodicity of Individual Objects} \label{sec:detect_period}

The most direct way to confirm the asteroid belt hypothesis is
to detect one or more individual objects by identifying features in the 
power spectrum that are statistically significant.
The power spectral density formed from the $N$-point discrete Fourier transform
$\tilde X_k$ of discretely sampled data is 
\be
P_k =  \frac{T}{ N^2} \vert \tilde X_k \vert^2, 
\ee
where $T$ is the observing span. 

Including
additive white noise with the timing perturbations from a collection of $N_a$ non-interacting asteroids,  the power spectrum has an ensemble average
\be
\label{eqn:ps_signal}
\langle P_k  \rangle = \langle P_{k,a} \rangle + \langle P_{k,w} \rangle,
\ee
where $\langle P_{k, w} \rangle$ is the mean white-noise
spectrum, 
\be
\langle P_{k,w} \rangle = \frac{T \sigma_w^2}{N},
\ee
where $\sigma_w$ is the rms white noise error.

The mean asteroid-induced spectrum is
\be
\label{eqn:ps_asteroid}
\langle P_{k,a} \rangle &=& 
 \frac{T}{2}\sum_{a=1}^{N_a} \sigma_{2}^2 
		\left[
			{\rm sincd}^2\pi(k/N-f_a\Delta t)
			\right.
			\nonumber \\ 
			&+& \left. {\rm sincd}^2\pi(k/N+f_a\Delta t )
		\right],
\label{eq:Pk}
\ee
where we have used the discrete ``sinc'' function defined as
${\rm sincd}(x) = \sin Nx / N\sin x$, which is unity for $x=0$. 
In the null hypothesis where there is no signal, the power spectral estimates
have a $\chi^2$ distribution with two degrees of freedom.  Defining
a normalized spectrum to be
\be
S_k = \frac{P_k} { \langle P_{k, w} \rangle},
\ee
for the null case, the probability that any single value
$S_k$ exceeds a value $S$ is
\be
P\{S_k > S\} = \exp(-S). 
\ee
When $M=N/2 \gg 1$ spectral amplitudes are
tested and we require  the mean number of false positives to be less than one,
\be
M \exp(-S) < 1,
\ee
we obtain
\be
S > \ln M.
\ee
More generally, the probability that one or more out of $M$ spectral
values exceed $S$ is the false-alarm
probability $p_{M}$
\be
p_{M} = P_{\ge 1}(>S) = 1 - [1 - \exp(-S)]^M.
\ee
An asteroid signal would need to be larger than $S$ for it to not be
considered a false-positive; equivalently, we can say that the null hypothesis
can be rejected with probability $1-p_{M}$.  This implies that
\be
S >  -\ln \left [ 1 - (1-p_M)^{1/M}\right]
\ee
or, for $p_M\ll 1$,
\be
S >  \ln M/p_M.
\ee

\subsubsubsection{Detecting Periodicities in a Sparse Asteroid Belt}
We first consider a sparse asteroid disk that contains only a few objects
that produce spectral lines in Equation (\ref{eq:Pk}) that do not overlap and thus
potentially can be identified.   In this case, detectability is limited solely by the
additive white noise.
For an asteroid signal that maximizes at $k_a$, this requirement becomes
approximately
\be
\label{eqn:ps_ineq}
P_{k_a}  >  \langle P_{k,{\rm off}} \rangle \ln (M/p_M).
\ee
Substituting from Equation (\ref{eqn:ps_asteroid}) and using $P_{k,\rm off} = T \sigma_w^2/N$
the asteroid signal must satisfy
\be
\sigma_2 > 
	\sigma_w\left (\frac{2}{{N}}\right)^{1/2} 
	\left[\ln(\frac{M}{p_M})-1 \right]^{1/2},
\ee
where $\sigma_2$ is defined in Equation \ref{eqn:pert_single_object}.

The implied minimum detectable asteroid mass is then
\be
m_{\rm min} &=& 
	\left(\frac{\sqrt 2 c M_{NS} }{a\sin i \,F(P_{\rm orb}, T) }\right)
	\left(\frac{ \sigma_w}{ \sqrt N}\right)
	\nonumber \\
	&&\quad\quad\quad \times \left(\ln \frac{M}{p_M} -1 \right)^{1/2}.
\ee
Substituting values appropriate for PSR B1937+21 including a neutron
star mass of 1.4~$M_{\odot}$,  using a false-alarm
probability $p_M = 1/M$ (i.e. a threshold corresponding to a single false alarm),
and $M=N/2$, the minimum detectable asteroid
mass is
\be
  m_{\rm min} &\approx& \frac{10^{-4.9}M_{\earth}}{\sin i\,F(P_{\rm orb}, T)} 
  	\left( \frac{\sigma_w}{\rm 100~ns} \right)
	\left(\frac{10^3}{N}\right)^{1/2}
	 \nonumber \\
  && \times  
   \left( \frac{1~\rm yr}{P_{\rm orb}}\right)^{-2/3}
  \left( \frac{2\ln N -1}{2\ln 10^3 - 1} \right)^{1/2}.
\ee

Individual objects of approximately one third the mass of the main belt asteroid Ceres can be detected in the Fourier transform if they are orbiting the pulsar at  $1$~AU, and only white noise is present in the observations.

Resolving spectral lines associated with a large number of objects is more difficult, however.
If the asteroids comprise a marginally stable and maximally packed belt with separations $\Delta r /r  \approx 10 (m/3 M_{\rm NS})^{1/3}$, the frequency resolution $\delta f= T^{-1}$ must be smaller than the separation of spectral lines produced by neighboring asteroids
$\delta f/f = (3/2) \Delta r/r$, which, for a $1.4 M_\sun$ neutron star works out to
\be
\frac{T}{P_{\rm orb}} \gtrsim 161 \left( \frac{10^{-4} M_\Earth}{m}\right)^{1/3}.
\ee
For data spans of 10 to 20 years, the frequency resolution will be sufficient
to resolve the spectral line from an individual asteroid only if the disk is much less populated than a maximally packed disk or if the asteroid has a mass much larger
than the average.  

\subsubsubsection{Detecting  Periodicities in a Dense Asteroid Belt}

Even if the cumulative effect of the asteroid belt produces a red noise-like signature, it may  still be possible to isolate the periodicity from the largest object or objects in the disk.
Using the PSD in Figure \ref{fig:PS_B1937} as a basis, we model the total noise background from which an individual asteroid needs to be distinguished
by a combination of white noise and a steep red component:

\be
\label{eqn:ps_noise}
P_{k,\rm off}(k=fT) &&=  
\nonumber \\
&&
\!\!\!\!\!\!\!\!\!\!\!\!
\!\!\!\!\!\!\!\!\!\!\!\!
0.015~{\rm \mu s^{2}~yr}  \left( \frac{f}{\rm 1 yr^{-1}}\right)^{-5} + 0.015~{\rm \mu s}^2~{\rm yr}.
\ee

\begin{figure}[!ht]
\begin{center} 
\includegraphics[scale=0.5]{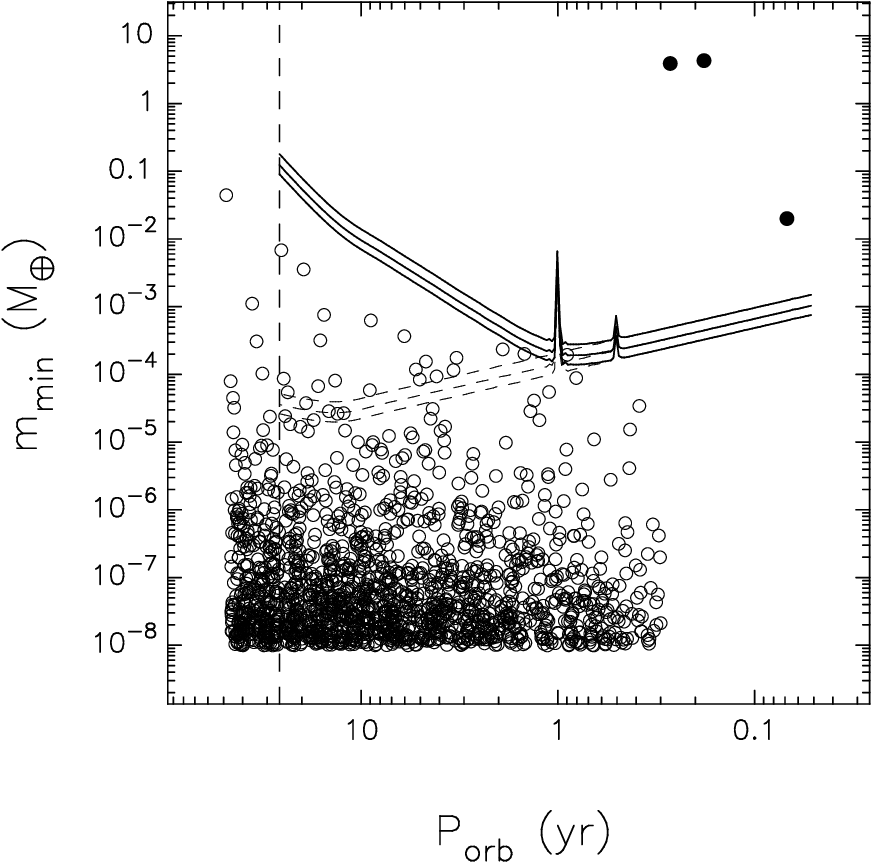}
\caption{ \footnotesize \label{fig:min_mass} Minimum detectable mass of an asteroidal  companion to PSR~B1937$+$21 for a dense asteroid belt.  The solid lines show the sensitivity assuming that the residuals contain red noise of the form presented in Equation (\ref{eqn:ps_noise}), which is a model consistent with the observed red noise. From bottom to top, the lines represent false alarm probabilities of $p=0.33$, $0.05$, and $10^{-4.2}$, which roughly correspond to $1-\sigma$, $2-\sigma$, and $5-\sigma$ limits.  The dashed curves show the sensitivity assuming that the only source of noise is white noise for the same set of false alarm probabilities. The filled circles denote the three planets orbiting the millisecond pulsar PSR~B1257$+$12 \cite[][]{2003ApJ...591L.147K},  which all would have been detected if present around PSR~B1937$+$21.
The open circles denote show the masses of the objects the best-fitting configurations displayed in panels $d-f$ of Figure \ref{fig:best_config}.
The vertical dashed line shows the data span length.  
The change in slope and loss of sensitivity for $P_{\rm orb} > 20$~yr is associated with the polynomial fit for pulsar spin down.  The narrow spikes at 0.5 and 1~yr  correspond to the
fit for parallax and proper motion, respectively. 
}
  \end{center}
\end{figure}

The minimum detectable mass can be found by substituting Equations (\ref{eqn:ps_noise}) and (\ref{eqn:ps_asteroid}) into Equation (\ref{eqn:ps_ineq}) and solving for $m$.
In Figure \ref{fig:min_mass}  we show a plot of the minimum detectable mass as a function of orbital period $P_{\rm orb}$.
When  $P_{\rm orb} \gg 1$~yr the sensitivity is poor because of large levels of red noise.    When $P_{\rm orb} \ll 1$~yr,  white noise dominates and sensitivity is poor because the reflex motion per unit mass is diminished.  Over a wide range of orbital periods we can rule out any single object with a mass $< 0.05 M_\earth$.  
In Figure  \ref{fig:min_mass} we also show the distribution of objects in mass and frequency from one of our best simulations.  As expected, none of the objects exceed the threshold for detectability in a search for their periodicity.
  
\subsubsection{Stationarity in Time Series} 

 If individual asteroids cannot be discriminated, the statistics of their collective behavior can be used to distinguish the asteroid model from intrinsic spin noise. 
Spin noise is manifested as non-stationarity\footnote{We note for some canonical
pulsars, it appears that there is a sub-dominant quasiperiodic source of RN attributed to magnetospheric activity \cite[][]{2010Sci...329..408L}.   We refer the reader to Section \ref{sec:TN_mag_activity} for further discussion.} in the residuals  (\citealt{1985ApJS...59..343C}; \citealt{sc2010}) that can be modeled as random walks in spin phase, spin frequency and frequency derivative, and other similar processes having power spectra that are power-law in form.   
For these processes respectively, the rms residuals scale as $\sigma_{{\rm RN},2}(T) \propto T^\gamma$, with $\gamma=1/2, 3/2,5/2$.
 Most CPs show consistency with the frequency and frequency derivative spin noise,  a superposition of the two, or random walks combined with occasional discrete jumps \cite[][]{1985ApJS...59..343C,1995MNRAS.277.1033D}.   
 More recently \cite{sc2010} found that for  rotation-powered pulsars   $\sigma_{{\rm RN},2}(T) \propto T^{2.0\pm 0.2}$.  The dependence on the data span $T$
is a clear signature of non-stationary statistics.   
 
In the asteroid model, for time scales shorter than the largest orbital period 
(i.e, $T <P_{\rm orb, max}$),
$\sigma_{{\rm RN},2}$ increases with $T$ because larger fractions of the 
perturbations associated with the longest period objects are manifested 
in the post fit residuals.    
However, 
long term stationarity is expected from asteroid belts for $T \gg P_{\rm orb, max}$.
 The transition from non-stationary to stationary statistics can be explored by applying a variety of diagnostics to the residual TOAs, including  structure functions, spectral estimators, and $\sigma_{{\rm RN},2}(T)$ itself.  
If non-stationarity persists for long periods of time (much greater than $25$~years), either an asteroid belt model would need to be developed that includes objects in very wide, stable orbits, or the asteroid belt origin for observed TOA variations could be ruled out.

\subsubsection{No Asteroids Inside the Rock Line}  

As mentioned earlier, the spin-down luminosity of PSR B1937$+$21 is large enough 
to heat asteroids to 
the large temperatures  
needed to evaporate them
(c.f., Equation~\ref{eqn:disk_temp_radius}).
For PSR B1937$+$21, asteroids composed of rocky material
would not be expected inside $0.5$~AU ($P_{\rm orb} \approx 0.4$~yr), assuming a melting temperature of $\approx 1500$~K and complete absorption of the pulsar spin down luminosity ($\epsilon=1$) and isotropic beaming ($g_b=1$).  
This rock line is analogous to the water/ice line discussed in planetary and exoplanetary science.  
Asteroids composed of ice and other refractory materials would not be expected inside a radius of $16$~AU ($P_{\rm orb} \approx 14$~yr), assuming a melting temperature of $300$~K and complete absorption of 
isotropically beamed pulsar spin down luminosity.
The absence of asteroids inside the rock line would be manifested as a paucity of spectral power at fluctuation frequencies $f > 1/P_{\rm orb,rock}$, where $P_{\rm orb,rock}$ is the orbital period at the rock line. The spectra in Figures~ \ref{fig:TN_B1937}  and 
\ref{fig:best_config} are consistent with this picture.  In order to detect the presence of red noise in this region, it would be necessary to reduce radiometer noise, by observing with a higher gain system, a wider bandwidth, or a longer integration time.   


\newpage 
\subsubsection{Differences in Red Noise between Isolated and Binary MSPs}


Relative to binary MSPs, the processes that form isolated MSPs may provide initially more massive and extended disks of material.  This is plausible  because isolated MSPs have presumably completely disrupted their companion and polluted their environment with a larger mass of material with a wider range of angular momentum  \cite[][]{2007ApJ...666.1232C}.
    In this scenario, as a population, isolated MSPs would possess larger and more radially extended disks, which would be expected to have larger objects in wider orbits and would have larger levels of red noise, as expressed in Equation (\ref{eqn:total_pert}).

It should be noted that binary systems themselves can potentially possess stable circumbinary disks because only the 
inner regions of the disk are strongly  torqued by the binary pair.     
Indeed, long lived debris disks have been observed around main sequence binary systems  \cite[][]{2007ApJ...658.1289T}.
In numerical simulations of late-stage planet formation,  \cite{2006Icar..185....1Q} found that planet formation surrounding compact binary star systems was statistically identical to planet formation around isolated stars, provided the binary systems were compact and in circular orbits.    \cite{1999AJ....117..621H} examined the stability of single planet orbits around binary systems, in which the binary has a wide range of eccentricities and mass ratios.   
Through numerical simulation, they derived an empirical relationship for the innermost stable orbit as a function of semi-major axis, eccentricity, and mass ratio of the binary.  For mass ratios between 1:1 and 9:1, and systems with eccentricities between $0$ and $0.7$, they found that the planets were stable with orbits that were a factor of four greater than the semi-major axis of the binary.  However, simulations were conducted for only $10^4$ orbits, which is more than $5$ orders of magnitude shorter than the time  required for MSP systems.  
Because NS-WD binary orbits are typically very compact ($P_{\rm orb} ~$few~hr to $80$~d), asteroids orbiting at $> 1$~AU are not strongly affected by the inner binary pair. 

Disks in principle could persist around double neutron star (NS-NS) systems as well. However, pulsars in NS-NS binaries do not exhibit the timing precision that MSPs, even without asteroidal effects.
Firstly, timing residuals from NS-NS binaries are larger because pulse widths and there are fewer pulses
per unit time interval, so pulse-phase jitter is larger.   
Secondly  there appears to be timing noise and intrinsic spin instabilities in some of the objects \cite[][]{2010ApJ...722.1030W}.  
Therefore the perturbations of an  asteroid belt that is allowed
around B1937+21 would be masked by the other timing effects. 
An asteroid belt would have to be accordingly more massive or have a small number of more massive individual asteroids for there to be a competing effect from asteroids.



At present, there is  sparse and perhaps only circumstantial
evidence that MSPs with massive companions are 
intrinsically more spin stable than isolated MSPs.
Indeed,  neither the PSR~B1257$+$12 system nor PSR~B1937$+$21 have stellar companions.   
By contrast, 
the  MSPs with the best timing stability have WD companions 
(PSRs J0437$-$4715, J1713$+$0747, and J1909$-$3744;  \citealt{2009MNRAS.400..951V}; \citealt{2012arXiv1201.6641D}).  

However there are a number of isolated MSPs that do not show the cubic signature 
of strong RN at the current levels of timing precision 
\cite[e.g., PSR~J2124$-$3358,][]{2009MNRAS.400..951V}.  
 Evidence for RN in more isolated MSPs would be necessary to show a definitive difference.  Even then, careful analysis would be required to conclusively identify that the correlation is due to an asteroid-belt origin and not  other plausible origins for red noise. 
 
\subsubsection{Asteroid Belt Size and Composition Correlated with MSP Age}
 
Asteroid belts may evolve slowly over the age of the pulsar.
Radiation,  gravitational migration, and occasional collisions will
cause objects to migrate inward toward the pulsar, particularly if the disk is more dense than minimally required. 
As material migrates inward, the reflex motion of the NS is diminished. 
Disks would be expected to be more compact, and (potentially much) smaller in mass around the oldest MSPs.

\subsubsection{Residuals Uncorrelated with Magnetospheric Activity}  \label{sec:TN_mag_activity}

\cite{2010Sci...329..408L} found that some of the RN in some
 CPs is associated with changes in state between discrete values of the
  spin down rate, $\dot{\nu}$, and that the states are associated with different average profile
   shapes. 
While there is no evidence for this process occurring in PSR~B1937$+$21,
if {\em all} of the RN was associated with this type of behavior, an asteroid belt origin would be ruled out, because reflex motion would neither cause transitions between discrete states of $\dot{\nu}$ nor cause the temporal variability of pulse profiles.  
   As discussed earlier,  the level of RN for this pulsar is consistent with what is observed in CPs \cite[][]{sc2010}.   
   However, given  pulsar to pulsar variability in the levels of RN in CPs, some fraction of the observed RN in this PSR~B1937$+$21 may still be associated with reflex motion.  
 
\subsubsection{Variations in Dispersion Measure}

 Another possibility is to 
search for variations in dispersion measure (DM) due to evaporation of asteroidal material by the pulsar's relativistic wind.  Whether DM variations are detectable depends on the evaporation rate and on how fast thermal gas is entrained in the relativistic flow.   Relativistic gas will produce a smaller timing perturbation than thermal gas with the same density. Some
fraction of the total measured DM from B1937+21 could be associated with thermal gas
near the pulsar, most likely just outside the bow shock that, for this object would be 
$~0.04$~pc from the pulsar using appropriate scaling laws
\citep[e.g.][]{ 2002ApJ...575..407C} for an ambient density of 1~H atom~cm$^{-3}$ and
a nominal three-dimensional space velocity of 100~km~s$^{-1}$.  However,  evaporation events much closer
to the pulsar might produce short-lived increases in DM,  but whether these occur or not
is related to the location of the rock line, as discussed above, and whether small asteroids
migrate to lower radii where they evaporate more quickly.     
Regardless,  variations in DM could provide a handle on circumpulsar debris.   Any timing
variations would need to be distinguished from interstellar variations, which produce
 $\approx 2~\mu$s at 1.4~GHz over time scales of years  \cite[][]{2006ApJ...645..303R}, and $\approx 100$ ns variations over time scales of $10$~days \cite[][]{2007MNRAS.378..493Y}. 
 DM variations from B1937+21 are consistent with an interstellar  origin because they
 conform to the expectations for  a Kolmogorov medium combined with discrete structures. Together  these produce flux-density variations that are anti-correlated with timing variations
 \cite[][]{1998A&A...334.1068L}.  Flux density variations would not be expected from
 evaporation events near the pulsar that conform to the observed anti-correlation, which is caused by refraction in a cold plasma.   
 While DM-induced timing variations are  larger at lower frequencies, scaling
 proportional to $\nu^{-2}$, multi-path propagation
 delays associated with  the ISM scale even more strongly ($\propto \nu^{-4}$) and they can  limit the timing precision necessary to determine the dispersion measure at each epoch \cite[][]{1990ApJ...364..123F}.   Thus, while searching for DM variations from circumpulsar material may be 
 difficult for PSR~B1937+21, other MSPs, especially those that are nearer the solar system and 
thus have smaller ISM contributions, may be fruitful targets for low-frequency arrays, such
as LOFAR  \cite[][]{2011A&A...530A..80S}.


\subsection{Other Tests Asteroid Belts around Pulsars}

\subsubsection{ Infrared Emission from Debris} 

Any debris  around neutron stars would be passively heated by the neutron star's thermal and non-thermal emission  to temperatures of hundreds of Kelvin, radiating predominantly in the infrared.  
For  debris to be detectable, the disk would be required to have a large effective area.  
 Infrared observations of other pulsars have placed limits of many Earth masses on circumpulsar dust disks \cite[][]{2004AJ....128..842L,2006ApJ...646.1038B}. Because of the large distance to PSR~B1937$+$21 \cite[$ 5^{+2}_{-1}$~kpc,][]{2012ApJ...755...39V},
 tens to hundreds of hours of telescope time with a JWST-like telescope would be required to detect an asteroid-belt like disk \cite[][]{2008ApJ...682.1152C}.
 The lack of a population of small radiating particles with a cumulative large effective area (removed from the system due to Poynting-Robertson drag) further inhibits the detection of the disk. 
 In addition, definitive association of infrared emission would be severely hindered by confusion with coincident infrared sources because the pulsar is located in the Galactic plane.

\subsubsection{Reflection of Radio Emission  off of Asteroids}  

If the pulsar beam intersects the asteroid belt, radio emission will reflect off of any material with size greater than the wavelength of the radio emission \cite[][]{1993ASPC...36..321P,2008ApJ...682.1152C}.
Reflected radio emission can be distinguished from magnetospheric emission through polarimetric observations because
reflected emission would be depolarized by  scattering off of the rough asteroid surfaces, while the magnetospheric emission would show substantial levels of polarization.
Detection of this emission is challenging with single-dish instrumentation because the faint off-pulse emission cannot be distinguished from receiver noise and other sources  within the comparatively larger beam.
In the past, radio interferometers have not been capable of producing the fast-dump imaging necessary to distinguish pulsed point source emission from non-pulsed reflection emission.  
However,  new instrumentation at the Australia Telescope Compact Array, Giant Metrewave Radio Telescope, Jansky Very Large Array,  MeerKAT (under construction), and  the planned Square Kilometre Array will be capable of fast-dump imaging and it will be possible to identify off pulse emission.  
Recently, using the Giant Metrewave Radio Telescope,  off-pulse emission has been associated with two CPs \cite[][]{2011ApJ...728..157B, 2012arXiv1208.6364B}.   However, this emission has been attributed with the pulsars' magnetospheres.

\section{Implications:  Timing Precision and Gravitational Wave Detection} \label{sec:implications_timing_precision}

If MSP timing precision is improved further it will be possible to detect gravitational waves (GWs) by observing time of arrival (TOA) variations that are correlated among a set of MSPs comprising a pulsar timing array \cite[PTA,][]{1979ApJ...234.1100D,1990ApJ...361..300F,2005ApJ...625L.123J}.   
The most plausible source of GWs is a stochastic background of gravitational  waves (GWB) associated with massive black hole binaries \cite[e.g.,][]{2008MNRAS.390..192S}. 

There are three ways to improve the sensitivity of a PTA to gravitational waves:  observe a larger number of pulsars, make more precise estimates of TOAs, or observe over a longer timing baseline. 
As long as timing error is dominated by the finite signal to noise ratio of the observations (i.e., radiometer noise),  timing precision  can be improved by observing the pulsars more frequently, with longer integration times, wider bandwidths, and higher gain radio telescopes.   However there are many reasons to expect limitations to these improvements (\citealt{2011MNRAS.418.1258O}; \citealt{2012arXiv1210.7021S}), including the presence of asteroid belts around MSPs.

In the previous sections, we have demonstrated that MSP systems that contain a population of large, rocky objects possess poorer timing residuals than those that do not. 
Assuming the disks consist of more than a few objects it is nearly impossible to mitigate the effects of these disks in residual TOAs.  
Most obviously, orbital noise cannot be reduced by simply performing more sensitive observations, e.g., by using more sensitive backend instrumentation or a higher gain telescope. 
Secondly and more alarmingly, the signature of orbital noise is similar to that imparted by the GWB on pulsar TOAs.  
As a result, for a given rms signal strength, the tolerance is much lower for orbital noise than white noise sources: reflex motion inducing $\approx 20$~ns rms variations over $5$ to $10$~yr will significantly reduce the sensitivity of a PTA to a GWB \cite[][]{sc2010}.

In the presence of red noise, there are two ways to improve PTA sensitivity to GWs.   A modest increase in sensitivity can be achieved by increasing observing throughput, i.e, by observing some combination of higher cadence and longer per-epoch observations. 
The only way to significantly increase sensitivity is to incorporate additional MSPs with comparable (or superior) stability into PTA observations, such as pulsars with lower levels of intrinsic red noise or smaller (or absent) asteroid disks. 
Indeed there are a handful of pulsars that are sensitive to the GWB within its predicted amplitude range and are absent of strong red noise \cite[][]{2012arXiv1201.6641D,2012arXiv1210.6130M}.  For the purposes of detection, it is imperative to identify other systems that show similar stability.   These implications are discussed in detail in \cite{2011arXiv1106.4047C}.



\section{Conclusions}

We have shown that asteroid belts can form and persist around MSPs, and these systems can affect the timing precision at the ns~to~$\mu$s level.  
The residual RN in observations of PSR B1937$+$21 is consistent with the presence of an asteroid belt of mass $\approx 0.05 M_\earth$.   
The mass and distribution of objects of objects in the disk is constrained by the timing variations.  However, additional constraints can be placed because the disk must be dynamically stable and asteroids need need to withstand the strong radiation field emitted by the pulsar.   In Table \ref{tab:assumptions} we list constraints on the disk configuration  derived in this paper, including the minimum and maximum mass, the mass and spatial distribution of the asteroids, and the number of objects.

\begin{deluxetable*}{llcl}
\tabletypesize{\footnotesize} \tablecolumns{4}
 \tablecaption{Constraints on Disk Configuration \label{tab:assumptions}}
\tablehead{   \colhead{Constraint} & \colhead{Mechanism} &\colhead{Reference} & \colhead{Comment}}   
\startdata
 Mass Distribution (obs) &  Reflex Motion  & Eqs. (4)-(10), (20) & Dependent on asteroid mass distribution (factor $\zeta$ in Eq. 8)  \\
 Maximum Asteroid Mass (obs)&  Reflex Motion & Figure 5 & Constrained by power spectral analysis to   $ M_{\rm max} \approx 0.05~M_\earth$\\ \hline 
Minimum Asteroid Size & Yarkovsky Migration &  Eq. (23) &  $R_{\min} \approx  5$~km, $M_{\rm min} \approx 10^{-10} M_{\earth}$ at $1$~AU \\ 
  Minimum Orbital Radius &  Radiative Heating & Eq. (13) &  $a_{\rm min} \approx 0.5$~AU, $P_{\rm orb, min} \approx 0.3$~yr if $T_{\rm melt} = 1500$~K  \\
Orbital Separation& Dynamical Stability  & Eqs. (16)-(17) &  $\Delta r > 10 r_H$ (extrapolated from Chambers et al. 1996)   \\
Maximum Number of Objects &  Dynamical Stability & Eqs.(18)-(20) & Dependent on minimum orbital separation 
 \enddata
  \tablecomments{Constraints on asteroid belt configurations derived in this paper.  For each constraint we list the mechanism that provides the constraint, the relevant equation or figures in the text, and a brief description of the magnitude of the constraint or assumptions required to set its magnitude.  Constraints that are derived from analysis of the time series are listed as (obs).  }
 \end{deluxetable*}

The existence of planets around two pulsars, a debris disk around another, and a possible asteroid belt around PSR~B1937$+$21 have several implications.  
Protoplanetary disks around pulsars must cover a wide range in density and total mass in order to have planets form in some cases but not in others.  
Disks of lower mass may exist around other MSPs, but not have been detected because of insufficient timing precision.

Although it is difficult to test the asteroid model, it may be equally difficult
to argue that absolutely no debris would be left in the system once accretion
driven spinup and evaporation of the companion were complete.
Confirmation of the orbital noise interpretation for PSR~B1937$+$21  may rely on finding planets and asteroids around other pulsars and demonstrating that they are a frequent outcome of processes that lead to recycled pulsars that presently do not have  a single large companion.

If disks are populated by more than a  few objects, it is difficult to identify individual objects and correct for their TOA perturbations.
Unless PSR~B1937$+$21 is unique, high precision timing observations of other MSPs may show unmitigatable, correlated orbital noise associated with circumpulsar  disks. More MSPs need to be investigated in detail to search for red noise signatures consistent with astroid disks. 

\acknowledgements

We thank the referee for  comments that improved the quality of this paper.
This work was supported by the NSF through grant AST-0807151 and the Research Experience for Undergraduates Program.  This work was also supported by NAIC, when it was operated by Cornell University under a cooperative agreement with the NSF. 
Part of this research was carried out at the Jet Propulsion Laboratory, California Institute of Technology, under a contract with the National Aeronautics and Space Administration.
This work is partially based on observations with the 100-m telescope of the
Max-Planck-Institut f\"ur Radioastronomie (MPIfR) at Effelsberg, Germany.
The Westerbork Synthesis Radio Telescope is operated by ASTRON
(Netherlands Foundation for Research in Astronomy) with support from
the Netherlands Foundation for Scientific Research (NWO).
The Nan\c cay Radio Observatory is operated by the Paris Observatory,
associated to the French Centre National de la Recherche Scientifique
(CNRS). The Nan\c cay Observatory also gratefully acknowledges the
financial support of the Region Centre in France.
This work made use of NASA's ADS System.

\appendix
\section{Observations of B1937$+$21} \label{sec:app_1937_obs}

In this appendix, we outline the methods used to form the residual times 
series displayed in Figure \ref{fig:TN_B1937} and the power spectrum 
in Figure \ref{fig:PS_B1937}  that are the basis for the analysis 
in this paper.
Observations are from the Arecibo, Effelsberg, Nan\c{c}ay, and Westerbork 
telescopes.  
In Table \ref{tab:1937_obs}, for each telescope  we list the  MJD range of observations,  observing span $T$, nominal observing frequency $\nu$, observing backend, the typical uncertainty in TOA, and the  total number of TOAs $N_{\rm TOA}$.  In this table, we also give a reference that describes the instrumentation.     
All of the observations from the Effelsberg and Nan\c{c}ay telescopes were conducted in bands close to $1400$~MHz.   
Observations at Arecibo were conducted in two widely separated frequency bands, while observations at Westerbork were conducted in three widely separated bands. 
All observations were corrected for known offsets in the observatory clocks, including a clock offset recently found in Westerbork data (G.~Janssen et al., in preparation). 
The TEMPO2 package \cite[][]{2006MNRAS.369..655H} was used to analyze the TOAs.
In the next two subsections, we describe how the timing solution used to for this analysis was made. 

\subsection{Correcting for Propagation Effects} 

It is well known that it is necessary to correct for interstellar propagation to achieve the best timing precision.  
Because the refractive index of the interstellar medium is frequency dependent, propagating radio waves at different frequencies arrive at different times and travel along different paths.    
This is manifested as a number of perturbations of the observed signal relative to the emitted signal
\citep[e.g.][]{1990ApJ...364..123F}.
At  minimum it is necessary to correct for the group delay associated with the total electron content of the line of sight, the latter referred to as the  dispersion measure (DM) in the parlance of pulsar astronomy.

DM variations in the Arecibo data were corrected using the published DM time series that accompany the dual frequency TOAs \cite[]{1994ApJ...428..713K}.
For the earliest Nan\c{c}ay observations (${\rm MJD} < 52579$), 
DM variations were corrected using published DM values 
\citep{2006ApJ...645..303R} that were  partially derived from the 
observations presented here.   For later observations,  the multi-frequency Westerbork observations were used to model DM variations using a ninth-order polynomial  \cite[][]{2011MNRAS.414.3134L}.
While the DM variations are significant, their
contribution to the timing error is sub-dominant to the achromatic red noise in the post fit residuals. This is because the largest contribution to DM variations is a linear trend associated with a DM gradient, which is absorbed into the fitting of spin frequency in single-frequency observations. 
  

\subsection{Determining Observatory and Receiver Backend Offsets}


It is essential to account for phase offsets between telescopes and backend instrumentation that are commonly referred to as {\em jumps}.  Phase offsets were measured relative to the high quality Westerbork observations using only overlapping data between different data sets and the technique outlined in \cite{2008A&A...490..753J}. 
  Errors in these offsets (possibly exacerbated by DM uncertainties) will introduce 
red noise into the residuals.  However, this noise is sub-dominant to the observed red noise because it contains both a smaller rms amplitude (as estimated from the uncertainties in the jumps) and a shallower fluctuation spectrum because a sequence of imperfectly removed jumps will appear as a random walk in pulse phase, which has (approximately) a spectrum $\propto f^{-2}$. 

\subsection{Calculating Pulsar Timing Solution}

Using the measured receiver and backend offsets, the combined data set was used to determine the pulsar position and proper motion.   We did not attempt to measure pulsar parallax.
While fitting for astrometric terms, the red noise  was modeled using higher order frequency derivatives (up to and including $d^8 \nu/dt^8$). 
With the astrometric terms fixed, a solution including only $\nu$ and $d\nu/dt$ was produced that was used in the analysis presented in the main text.


\section{Calculating Maximum Entropy Power Spectra}\label{app:maxent}

We  use a maximum entropy spectral estimator \cite[][]{1163713} to form power spectra.    This type of spectral estimator was used because unlike a Fourier transform, it is not affected by spectral leakage.  This feature enables us to properly model the red signal observed in the residual TOAs of PSR~B1937$+$21 and produced by the simulations of asteroid belts.   

To form the maximum entropy power spectra, we first used cubic spline interpolation to evenly sample the observations onto a weekly grid. 
 The maximum entropy spectrum was then constructed using an order~$150$  autoregressive model of the interpolated data.   
A spectrum was calculated using $1000$ frequency bins spaced evenly between $0.038$~yr$^{-1} = 1/T$  and $10$~yr$^{-1}$.   
This number of bins provided oversampling sufficient to detect features in the power spectra associated with single dominant periodicities \cite[][]{Numerical-Recipes}.
  

\section{Calculating the Transmission Function
$F(P_{\rm orb}, T)$}\label{app:transfunc}

For objects with orbital periods $P_{\rm orb}$ much greater than the observing span $T$, a significant amount, but not all,  of the signature induced in the TOAs will be absorbed by the fit for pulsar spin down and astrometric terms.  We quantify  the amount of attenuation by a factor $F(P_{\rm orb},T)$.  In this section we describe the simulations that we used to determine the form of this factor appropriate for our observations.

We conducted a series of simulated TOAs, with observing cadence matching the actual observations.  In each simulation, the residual TOAs contained a single sinusoid with period $P_{\rm orb}$. 
To account for the TOA modeling process, we fit and removed a quadratic polynomial, annual, and semiannual sinusoids from the TOAs.
  We then calculated the ratio of prefit and postfit rms.   To produce an estimate of the ensemble average value of $F(P_{\rm orb},T)$, for each value of $P_{\rm orb}$, we injected the sinusoid a $100$ times with different, randomly chosen phases.  
  We calculated this for $200$ values of $P_{\rm orb}$ for values between $\approx 0.2$~yr and $210$~yr to produce the curve displayed in Figure \ref{fig:transfunc}.  We used a non-linear fitting routine to empirically find the functional form for the transmission function displayed in Equation (\ref{eqn:trans_func_relationship}). We numerically interpolated to produce the single-object sensitivity curves displayed in Figure \ref{fig:min_mass}.

    \begin{deluxetable}{lccccccc}
\tabletypesize{\footnotesize} \tablecolumns{8}
 \tablecaption{Observations used in this analysis\label{tab:1937_obs}}
\tablehead{ \colhead{Observatory}  & \colhead{MJD Range} &\colhead{$T$} & \colhead{Backend} & \colhead{$\nu$} & \colhead{$N_{\rm TOA}$} & \colhead{$\sigma_{\rm TOA}$}  &\colhead{Ref.}\\
\colhead{} & \colhead{} & \colhead{(yr)} & \colhead{}  & \colhead{(GHz)}  &\colhead{} & \colhead{(ns)} &\colhead{}}   
\startdata
Arecibo &   $46024-48974$ & $8.1$&  See (1) & $1410,2380$ & $420$ &$200$ &1\\
Nan\c{c}ay & $47517-53339$ & $14.9$ & VCO& $1410$ & $3200$ & $300$ &2 \\
Nan\c{c}ay & $53272-54730$ & $4.0$ & BON & $1370$ & $194$ &  $30$ &3\\
Effelsberg & $51557-54216$ & $7.3$ & EBPP & $1410$ & $161$ &$40$ &4 \\
Westerbork & $51566-55375$ & $10.4$ & PUMA & $840, 1380, 2280$ & $226$ &$100$ &5 \\
\hline
Total & $46024-55375$ & $25.8$ & \nodata & \nodata & $4201$ & \nodata
\enddata
  \tablecomments{We list the observatory, observation span, backend instrumentation,  typical observing frequency, typical TOA uncertainty $\sigma_{\rm TOA}$, and references that describe the instrumental setup.  References: (1) \cite{1994ApJ...428..713K}   ,  (2) \cite{1995A&A...296..169C}, (3) \cite{2009arXiv0911.1612C}, (4) \cite{2000ASPC..202...61L}, and (5) \cite{2002A&A...385..733V}.   }
  \end{deluxetable}



\begin{thebibliography}{70}
\expandafter\ifx\csname natexlab\endcsname\relax\def\natexlab#1{#1}\fi

\bibitem[Arzoumanian et al.(1999)]{1999ApJ...520..696A} Arzoumanian, Z., 
Cordes, J.~M., \& Wasserman, I.\ 1999, \apj, 520, 696 

\bibitem[{{Backer} {et~al.}(1993){Backer}, {Foster}, \&
  {Sallmen}}]{1993Natur.365..817B}
{Backer}, D.~C., {Foster}, R.~S., \& {Sallmen}, S. 1993, \nat, 365, 817

\bibitem[{Bailes} {et~al.}(2011)]{2011arXiv1108.5201B} Bailes, M., Bates, 
S.~D., Bhalerao, V., et al.\ 2011, Science, 333, 1717 

\bibitem[Basu et al.(2012)]{2012arXiv1208.6364B} Basu, R., Mitra, D., 
\& Athreya, R.\ 2012, \apj, 758, 91 


\bibitem[Basu et al.(2011)]{2011ApJ...728..157B} Basu, R., Athreya, R., 
\& Mitra, D.\ 2011, \apj, 728, 157 


\bibitem[{{Beloborodov} \& {Thompson}(2007)}]{2007ApJ...657..967B}
{Beloborodov}, A.~M., \& {Thompson}, C. 2007, \apj, 657, 967

\bibitem[{{Bryden} {et~al.}(2006){Bryden}, {Beichman}, {Rieke}, {Stansberry},
  {Stapelfeldt}, {Trilling}, {Turner}, \& {Wolszczan}}]{2006ApJ...646.1038B}
{Bryden}, G., {Beichman}, C.~A., {Rieke}, G.~H., {Stansberry}, J.~A.,
  {Stapelfeldt}, K.~R., {Trilling}, D.~E., {Turner}, N.~J., \& {Wolszczan}, A.
  2006, \apj, 646, 1038

\bibitem[{{Burns} {et~al.}(1979){Burns}, {Lamy}, \&
  {Soter}}]{1979Icar...40....1B}
{Burns}, J.~A., {Lamy}, P.~L., \& {Soter}, S. 1979, Icarus, 40, 1

\bibitem[{{Campana} {et~al.}(2011)}]{2011arXiv1112.0018C}  Campana, S., Lodato, 
G., D'Avanzo, P., et al.\ 2011, \nat, 480, 69 


\bibitem[{{Chambers} {et~al.}(1996){Chambers}, {Wetherill}, \&
  {Boss}}]{1996Icar..119..261C}
{Chambers}, J.~E., {Wetherill}, G.~W., \& {Boss}, A.~P. 1996, Icarus, 119, 261

\bibitem[{{Champion} {et~al.}(2010){Champion}, {Hobbs}, {Manchester},
  {Edwards}, {Backer}, {Bailes}, {Bhat}, {Burke-Spolaor}, {Coles}, {Demorest},
  {Ferdman}, {Folkner}, {Hotan}, {Kramer}, {Lommen}, {Nice}, {Purver},
  {Sarkissian}, {Stairs}, {van Straten}, {Verbiest}, \&
  {Yardley}}]{2010arXiv1008.3607C}
{Champion}, D.~J., {et~al.} 2010, \apjl, 720, L201

\bibitem[Chatterjee 
\& Cordes(2002)]{2002ApJ...575..407C} Chatterjee, S., \& Cordes, J.~M.\ 2002, \apj, 575, 407

\bibitem[{{Cognard} {et~al.}(1995){Cognard}, {Bourgois}, {Lestrade}, {Biraud},
  {Aubry}, {Darchy}, \& {Drouhin}}]{1995A&A...296..169C}
{Cognard}, I., {Bourgois}, G., {Lestrade}, J., {Biraud}, F., {Aubry}, D.,
  {Darchy}, B., \& {Drouhin}, J. 1995, \aap, 296, 169

\bibitem[{{Cognard} {et~al.}(2009){Cognard}, {Theureau}, {Desvignes}, \&
  {Ferdman}}]{2009arXiv0911.1612C}
{Cognard}, I., {Theureau}, G., {Desvignes}, G., \& {Ferdman}, R. 2009,
  arXiv:0911.1612

\bibitem[{{Cordes}(1993)}]{1993ASPC...36...43C}
{Cordes}, J.~M. 1993, in Astronomical Society of the Pacific Conference Series,
  Vol.~36, Planets Around Pulsars, ed. {J.~A.~Phillips, S.~E.~Thorsett, \&
  S.~R.~Kulkarni}, 43--60

\bibitem[{{Cordes} \& {Downs}(1985)}]{1985ApJS...59..343C}
{Cordes}, J.~M., \& {Downs}, G.~S. 1985, \apjs, 59, 343

\bibitem[{{Cordes} \& {Shannon}(2008)}]{2008ApJ...682.1152C}
{Cordes}, J.~M., \& {Shannon}, R.~M. 2008, \apj, 682, 1152


\bibitem[{{Cordes} \& {Shannon}(2012)}]{2011arXiv1106.4047C}
---. 2012, \apj, 750, 89

\bibitem[{{Cordes} {et~al.}(1990){Cordes}, {Wolszczan}, {Dewey}, {Blaskiewicz},
  \& {Stinebring}}]{1990ApJ...349..245C}
{Cordes}, J.~M., {Wolszczan}, A., {Dewey}, R.~J., {Blaskiewicz}, M., \&
  {Stinebring}, D.~R. 1990, \apj, 349, 245

\bibitem[{{Currie} \& {Hansen}(2007)}]{2007ApJ...666.1232C}
{Currie}, T., \& {Hansen}, B. 2007, \apj, 666, 1232

\bibitem[{{D'Alessandro} {et~al.}(1995){D'Alessandro}, {McCulloch}, {Hamilton},
  \& {Deshpande}}]{1995MNRAS.277.1033D}
{D'Alessandro}, F., {McCulloch}, P.~M., {Hamilton}, P.~A., \& {Deshpande},
  A.~A. 1995, \mnras, 277, 1033

\bibitem[{{Debes} \& {Sigurdsson}(2002)}]{2002ApJ...572..556D}
{Debes}, J.~H., \& {Sigurdsson}, S. 2002, \apj, 572, 556

\bibitem[{{Demorest} {et~al.}(2010){Demorest}, {Pennucci}, {Ransom}, {Roberts},
  \& {Hessels}}]{2010Natur.467.1081D}
{Demorest}, P.~B., {Pennucci}, T., {Ransom}, S.~M., {Roberts}, M.~S.~E., \&
  {Hessels}, J.~W.~T. 2010, \nat, 467, 1081

\bibitem[{{Demorest} {et~al.}(2013)}]{2012arXiv1201.6641D} Demorest, P.~B., 
Ferdman, R.~D., Gonzalez, M.~E., et al.\ 2013, \apj, 762, 94 


\bibitem[{{Detweiler}(1979)}]{1979ApJ...234.1100D}
{Detweiler}, S. 1979, \apj, 234, 1100

\bibitem[{{Dohnanyi}(1969)}]{1969JGR....74.2531D}
{Dohnanyi}, J.~W. 1969, \jgr, 74, 2531

\bibitem[{{Farihi} {et~al.}(2011){Farihi}, {Brinkworth}, {Gaensicke}, {Marsh},
  {Girven}, {Hoard}, {Klein}, \& {Koester}}]{2011arXiv1101.0158F}
 Farihi, J., Brinkworth,  C.~S., G{\"a}nsicke, B.~T., et al.\ 2011, \apjl, 728, L8 

\bibitem[{{Foster} \& {Backer}(1990)}]{1990ApJ...361..300F}
{Foster}, R.~S., \& {Backer}, D.~C. 1990, \apj, 361, 300

\bibitem[{{Foster} \& {Cordes}(1990)}]{1990ApJ...364..123F}
{Foster}, R.~S., \& {Cordes}, J.~M. 1990, \apj, 364, 123

\bibitem[{{Freire} {et~al.}(2012){Freire}, {Wex}, {Esposito-Far{\`e}se},
  {Verbiest}, {Bailes}, {Jacoby}, {Kramer}, {Stairs}, {Antoniadis}, \&
  {Janssen}}]{2012arXiv1205.1450F}
 Freire, P.~C.~C., Wex,  N., Esposito-Far{\`e}se, G., et al.\ 2012, \mnras, 423, 3328 

\bibitem[{{Fruchter} {et~al.}(1988){Fruchter}, {Stinebring}, \&
  {Taylor}}]{1988Natur.333..237F}
{Fruchter}, A.~S., {Stinebring}, D.~R., \& {Taylor}, J.~H. 1988, \nat, 333, 237

\bibitem[{{Goldreich} \& {Julian}(1969)}]{1969ApJ...157..869G}
{Goldreich}, P., \& {Julian}, W.~H. 1969, \apj, 157, 869

\bibitem[{{Hansen} {et~al.}(2009){Hansen}, {Shih}, \&
  {Currie}}]{2009ApJ...691..382H}
{Hansen}, B.~M.~S., {Shih}, H., \& {Currie}, T. 2009, \apj, 691, 382

\bibitem[{{Heng} \& {Tremaine}(2010)}]{2010MNRAS.401..867H}
{Heng}, K., \& {Tremaine}, S. 2010, \mnras, 401, 867

\bibitem[{{Hobbs} {et~al.}(2006){Hobbs}, {Edwards}, \&
  {Manchester}}]{2006MNRAS.369..655H}
{Hobbs}, G.~B., {Edwards}, R.~T., \& {Manchester}, R.~N. 2006, \mnras, 369, 655

\bibitem[{{Holman} \& {Wiegert}(1999)}]{1999AJ....117..621H}
{Holman}, M.~J., \& {Wiegert}, P.~A. 1999, \aj, 117, 621

\bibitem[{{Janssen} {et~al.}(2008){Janssen}, {Stappers}, {Kramer}, {Nice},
  {Jessner}, {Cognard}, \& {Purver}}]{2008A&A...490..753J}
{Janssen}, G.~H., {Stappers}, B.~W., {Kramer}, M., {Nice}, D.~J., {Jessner},
  A., {Cognard}, I., \& {Purver}, M.~B. 2008, \aap, 490, 753

\bibitem[{{Jenet} {et~al.}(2005){Jenet}, {Hobbs}, {Lee}, \&
  {Manchester}}]{2005ApJ...625L.123J}
{Jenet}, F.~A., {Hobbs}, G.~B., {Lee}, K.~J., \& {Manchester}, R.~N. 2005,
  \apjl, 625, L123

\bibitem[{{Johnston} {et~al.}(1992){Johnston}, {Manchester}, {Lyne}, {Bailes},
  {Kaspi}, {Qiao}, \& {D'Amico}}]{1992ApJ...387L..37J}
{Johnston}, S., {Manchester}, R.~N., {Lyne}, A.~G., {Bailes}, M., {Kaspi},
  V.~M., {Qiao}, G., \& {D'Amico}, N. 1992, \apjl, 387, L37

\bibitem[{{Jones}(1990)}]{1990MNRAS.246..364J}
{Jones}, P.~B. 1990, \mnras, 246, 364

\bibitem[{{Kaspi} {et~al.}(1994){Kaspi}, {Taylor}, \&
  {Ryba}}]{1994ApJ...428..713K}
{Kaspi}, V.~M., {Taylor}, J.~H., \& {Ryba}, M.~F. 1994, \apj, 428, 713

\bibitem[{{Koester} \& {Wilken}(2006)}]{2006A&A...453.1051K}
{Koester}, D., \& {Wilken}, D. 2006, \aap, 453, 1051

\bibitem[{{Konacki} \& {Wolszczan}(2003)}]{2003ApJ...591L.147K}
{Konacki}, M., \& {Wolszczan}, A. 2003, \apjl, 591, L147

\bibitem[{{Kramer} {et~al.}(2006){Kramer}, {Lyne}, {O'Brien}, {Jordan}, \&
  {Lorimer}}]{2006Sci...312..549K}
{Kramer}, M., {Lyne}, A.~G., {O'Brien}, J.~T., {Jordan}, C.~A., \& {Lorimer},
  D.~R. 2006, Science, 312, 549

\bibitem[{{Lange} {et~al.}(2000){Lange}, {Wex}, {Kramer}, {Doroshenko}, \&
  {Backer}}]{2000ASPC..202...61L}
{Lange}, C., {Wex}, N., {Kramer}, M., {Doroshenko}, O., \& {Backer}, D.~C.
  2000, in Astronomical Society of the Pacific Conference Series, Vol. 202, IAU
  Colloq. 177: Pulsar Astronomy - 2000 and Beyond, ed. {M.~Kramer, N.~Wex, \&
  R.~Wielebinski}, 61

\bibitem[{{Lazaridis} {et~al.}(2011){Lazaridis}, {Verbiest}, {Tauris},
  {Stappers}, {Kramer}, {Wex}, {Jessner}, {Cognard}, {Desvignes}, {Janssen},
  {Purver}, {Theureau}, {Bassa}, \& {Smits}}]{2011MNRAS.414.3134L}
{Lazaridis}, K., {et~al.} 2011, \mnras, 414, 3134

\bibitem[{{Lazio} \& {Fischer}(2004)}]{2004AJ....128..842L}
{Lazio}, T.~J.~W., \& {Fischer}, J. 2004, \aj, 128, 842

\bibitem[Lestrade et 
al.(1998)]{1998A&A...334.1068L} Lestrade, J.-F., Rickett, B.~J., \& Cognard, I.\ 1998, \aap, 334, 1068

\bibitem[{{Lin} {et~al.}(1991){Lin}, {Woosley}, \&
  {Bodenheimer}}]{1991Natur.353..827L}
{Lin}, D.~N.~C., {Woosley}, S.~E., \& {Bodenheimer}, P.~H. 1991, \nat, 353, 827

\bibitem[{{Lissauer}(1987)}]{1987Icar...69..249L}
{Lissauer}, J.~J. 1987, Icarus, 69, 249

\bibitem[{{Lissauer} \& {Stewart}(1993)}]{1993ASPC...36..217L}
{Lissauer}, J.~J., \& {Stewart}, G.~R. 1993, in Astronomical Society of the
  Pacific Conference Series, Vol.~36, Planets Around Pulsars, ed.
  {J.~A.~Phillips, S.~E.~Thorsett, \& S.~R.~Kulkarni}, 217--233

\bibitem[{{Lyne} {et~al.}(2010){Lyne}, {Hobbs}, {Kramer}, {Stairs}, \&
  {Stappers}}]{2010Sci...329..408L}
{Lyne}, A., {Hobbs}, G., {Kramer}, M., {Stairs}, I., \& {Stappers}, B. 2010,
  Science, 329, 408

\bibitem[Manchester et al.(2012)]{2012arXiv1210.6130M} Manchester, R.~N., 
Hobbs, G., Bailes, M., et al.\ 2012, arXiv:1210.6130 


 \bibitem[{{Miller} \& {Hamilton}(2001)}]{2001ApJ...550..863M}
{Miller}, M.~C., \& {Hamilton}, D.~P. 2001, \apj, 550, 863

\bibitem[{{Nakamura} \& {Piran}(1991)}]{1991ApJ...382L..81N}
{Nakamura}, T., \& {Piran}, T. 1991, \apjl, 382, L81

\bibitem[Os{\l}owski et al.(2011)]{2011MNRAS.418.1258O} Os{\l}owski, S., 
van Straten, W., Hobbs, G.~B., Bailes, M., 
\& Demorest, P.\ 2011, \mnras, 418, 1258 


\bibitem[{Papoulis(1981)}]{1163713}
Papoulis, A. 1981, Acoustics, Speech and Signal Processing, IEEE Transactions
  on, 29, 1176

\bibitem[{{Phillips}(1993)}]{1993ASPC...36..321P}
{Phillips}, J.~A. 1993, in ASP Conf. Ser. 36: Planets Around Pulsars, ed. J.~A.
  {Phillips}, S.~E. {Thorsett}, \& S.~R. {Kulkarni}, 321--325

\bibitem[{{Phinney} \& {Hansen}(1993)}]{1993ASPC...36..371P}
{Phinney}, E.~S., \& {Hansen}, B.~M.~S. 1993, in Astronomical Society of the
  Pacific Conference Series, Vol.~36, Planets Around Pulsars, ed.
  {J.~A.~Phillips, S.~E.~Thorsett, \& S.~R.~Kulkarni}, 371--390

\bibitem[{{Podsiadlowski}(1993)}]{1993ASPC...36..149P}
{Podsiadlowski}, P. 1993, in Astronomical Society of the Pacific Conference
  Series, Vol.~36, Planets Around Pulsars, ed. {J.~A.~Phillips, S.~E.~Thorsett,
  \& S.~R.~Kulkarni}, 149--165

\bibitem[{Press {et~al.}(1992)Press, Flannery, Teukolsky, \&
  Vetterling}]{Numerical-Recipes}
Press, W.~H., Flannery, B.~P., Teukolsky, S.~A., \& Vetterling, W.~T. 1992,
  Numerical Recipes: The Art of Scientific Computing, 2nd edn. (New York:
  Cambridge University Press)

\bibitem[{{Quintana} \& {Lissauer}(2006)}]{2006Icar..185....1Q}
{Quintana}, E.~V., \& {Lissauer}, J.~J. 2006, Icarus, 185, 1

\bibitem[{{Ramachandran} {et~al.}(2006){Ramachandran}, {Demorest}, {Backer},
  {Cognard}, \& {Lommen}}]{2006ApJ...645..303R}
{Ramachandran}, R., {Demorest}, P., {Backer}, D.~C., {Cognard}, I., \&
  {Lommen}, A. 2006, \apj, 645, 303

\bibitem[Roberts(2011)]{2011AIPC.1357..127R} Roberts, M.~S.~E.\ 2011, 
American Institute of Physics Conference Series, 1357, 127 

\bibitem[{{Ruden}(1993)}]{1993ASPC...36..197R}
{Ruden}, S.~P. 1993, in Astronomical Society of the Pacific Conference Series,
  Vol.~36, Planets Around Pulsars, ed. {J.~A.~Phillips, S.~E.~Thorsett, \&
  S.~R.~Kulkarni}, 197--215

\bibitem[{{Sesana} {et~al.}(2008){Sesana}, {Vecchio}, \&
  {Colacino}}]{2008MNRAS.390..192S}
{Sesana}, A., {Vecchio}, A., \& {Colacino}, C.~N. 2008, \mnras, 390, 192

\bibitem[{{Shannon} \& {Cordes}(2010)}]{sc2010}
{Shannon}, R.~M., \& {Cordes}, J.~M. 2010, \apj, 725, 1607

\bibitem[Shannon 
\& Cordes(2012)]{2012arXiv1210.7021S}  Shannon, R.~M., \& Cordes, J.~M.\ 2012, \apj, 761, 64 

\bibitem[{{Sigurdsson}(1993)}]{1993ApJ...415L..43S}
{Sigurdsson}, S. 1993, \apjl, 415, L43

\bibitem[{{Spitkovsky}(2011)}]{2011heep.conf..139S}
{Spitkovsky}, A. 2011, in High-Energy Emission from Pulsars and their Systems,
  ed. {D.~F.~Torres \& N.~Rea}, 139--158

\bibitem[Stappers et al.(2011)]{2011A&A...530A..80S} Stappers, B.~W., Hessels, J.~W.~T., Alexov, A., et al.\ 2011, \aap, 530, A80 

\bibitem[Tauris(2012)]{2012Sci...335..561T} Tauris, T.~M.\ 2012, Science, 
335, 561 

\bibitem[{{Terai} \& {Itoh}(2011)}]{2011PASJ...63..335T}
{Terai}, T., \& {Itoh}, Y. 2011, \pasj, 63, 335

\bibitem[{{Thorsett} {et~al.}(1999){Thorsett}, {Arzoumanian}, {Camilo}, \&
  {Lyne}}]{1999ApJ...523..763T}
{Thorsett}, S.~E., {Arzoumanian}, Z., {Camilo}, F., \& {Lyne}, A.~G. 1999,
  \apj, 523, 763

\bibitem[{{Trilling} {et~al.}(2007){Trilling}, {Stansberry}, {Stapelfeldt},
  {Rieke}, {Su}, {Gray}, {Corbally}, {Bryden}, {Chen}, {Boden}, \&
  {Beichman}}]{2007ApJ...658.1289T}
{Trilling}, D.~E., {et~al.} 2007, \apj, 658, 1289

\bibitem[{{Verbiest} {et~al.}(2009){Verbiest}, {Bailes}, {Coles}, {Hobbs}, {van
  Straten}, {Champion}, {Jenet}, {Manchester}, {Bhat}, {Sarkissian}, {Yardley},
  {Burke-Spolaor}, {Hotan}, \& {You}}]{2009MNRAS.400..951V}
{Verbiest}, J.~P.~W., {et~al.} 2009, \mnras, 400, 951


\bibitem[Verbiest et al.(2012)]{2012ApJ...755...39V} Verbiest, J.~P.~W., 
Weisberg, J.~M., Chael, A.~A., Lee, K.~J., 
\& Lorimer, D.~R.\ 2012, \apj, 755, 39 


\bibitem[{{Vo{\^u}te} {et~al.}(2002){Vo{\^u}te}, {Kouwenhoven}, {van Haren},
  {Langerak}, {Stappers}, {Driesens}, {Ramachandran}, \&
  {Beijaard}}]{2002A&A...385..733V}
{Vo{\^u}te}, J.~L.~L., {Kouwenhoven}, M.~L.~A., {van Haren}, P.~C., {Langerak},
  J.~J., {Stappers}, B.~W., {Driesens}, D., {Ramachandran}, R., \& {Beijaard},
  T.~D. 2002, \aap, 385, 733

\bibitem[{{Wang} {et~al.}(2006){Wang}, {Chakrabarty}, \&
  {Kaplan}}]{2006Natur.440..772W}
{Wang}, Z., {Chakrabarty}, D., \& {Kaplan}, D.~L. 2006, \nat, 440, 772

\bibitem[Weisberg et al.(2010)]{2010ApJ...722.1030W} Weisberg, J.~M., Nice, 
D.~J., \& Taylor, J.~H.\ 2010, \apj, 722, 1030 


\bibitem[Wex et al.(2000)]{2000ApJ...528..401W} Wex, N., Kalogera, V., 
\& Kramer, M.\ 2000, \apj, 528, 401

\bibitem[{{Wolszczan} \& {Frail}(1992)}]{1992Natur.355..145W}
{Wolszczan}, A., \& {Frail}, D.~A. 1992, \nat, 355, 145

\bibitem[{{You} {et~al.}(2007){You}, {Hobbs}, {Coles}, {Manchester}, {Edwards},
  {Bailes}, {Sarkissian}, {Verbiest}, {van Straten}, {Hotan}, {Ord}, {Jenet},
  {Bhat}, \& {Teoh}}]{2007MNRAS.378..493Y}
{You}, X.~P., {et~al.} 2007, \mnras, 378, 493

\end{thebibliography}

\end{document}